\documentclass[useAMS,usenatbib]{mn2e}

\usepackage{graphicx}
\usepackage{amsmath,amssymb}
\usepackage{mathrsfs}
\usepackage{times}    
\usepackage{colortbl}
\usepackage{nicefrac}

\definecolor{sgrey}{rgb}{0.44,0.50,0.56}
\definecolor{grey}{rgb}{0.75,0.75,0.75}

\def\mdrad{\ensuremath{\langle r_{\text{d}}\rangle}}
\def\rext{\ensuremath{r_{\text{ext}}}}
\def\rint{\ensuremath{r_{\text{int}}}}
\def\drad{\ensuremath{r_{\text{d}}}}
\def\taui{\ensuremath{\tau^{-1}_{\text{G}}}}
\def\CtHt{\ensuremath{\mbox{C}_2\mbox{H}_2}}
\def\rhod{\ensuremath{\rho_{\text{d}}}}
\def\teg{\ensuremath{T_{\text{g}}}}
\def\kms{\ensuremath{\,\mbox{km}\,\mbox{s}^{-1}}}

\def\CtoO{\ensuremath{\nicefrac{\varepsilon_{\text{C}}}{\varepsilon_{\text{O}}}}}
\def\drhog{\ensuremath{\nicefrac{\rho_{\text{d}}}{\rho}}}
\def\uinf{\ensuremath{u_{\infty}}}
\def\fcond{\ensuremath{f_{\text{cond}}}}
\def\vdri{\ensuremath{v_{\text{D}}}}
\def\ngrid{\ensuremath{n_{\text{g}}}}
\def\teff{\ensuremath{T_{\text{eff}}}}
\def\deltaup{\ensuremath{\Delta\,u_{\text{p}}}}
\def\mdotu{\ensuremath{\text{M}_{\sun}\,\text{yr}^{-1}}}
\def\sistd{\ensuremath{\sigma_{\text{s}}}}
\def\mmdot{\ensuremath{\langle\dot{M}\rangle}}
\def\mdmdot{\ensuremath{\langle\dot{M}_{\text{d}}\rangle}}
\def\muinf{\ensuremath{\langle u_{\infty}\rangle}}
\def\mfcond{\ensuremath{\langle f_{\text{cond}}\rangle}}
\def\mdrhog{\ensuremath{\langle\drhog\rangle}}

\def\mvdri{\ensuremath{\langle v_{\text{D}}\rangle}}
\def\mph{\ensuremath{_{i+\frac{1}{2}}}}
\def\mpm{\ensuremath{_{i-\frac{1}{2}}}}

\newcommand{\bfs}[1]{\textbf{\underline{#1}}}
\newcommand{\und}[1]{\underline{#1}}
\newcommand{\mref}{\multicolumn{1}{l}{\hspace*{0.35cm}\textsc{ref.}}}
\def\Rs{\ensuremath{R_{\star}}}
\def\rfluc{\ensuremath{\hat{r}}}
\def\pP{\ensuremath{P}}

\newcommand{\rSaHoa}{Paper~I}
\newcommand{\rSaHob}{Paper~II}
\newcommand{\rSaHoc}{Paper~III}
\newcommand{\rDD}{DD87}
\newcommand{\rDPSG}{D06}
\newcommand{\rWN}{WN86}
\newcommand{\rW}{W06}
\newcommand{\rKS}{KS97}

\title[Three-component modeling of C-rich AGB star winds. IV] 
  {Three-component modeling of C-rich AGB star winds\\
   IV. Revised interpretation with improved numerical descriptions}
\author[C.\ Sandin]{Christer Sandin\thanks{e-mail: CSandin@aip.de}\\%
   Astrophysikalisches Institut Potsdam, An der Sternwarte 16, D-14482 Potsdam, Germany}

\begin{document}

\date{Accepted 2007 October 25. Received 2007 October 23; in original form 2007 July 16}

\pagerange{\pageref{firstpage}--\pageref{lastpage}} \pubyear{2008}

\maketitle

\label{firstpage}

\begin{abstract}
Models describing dust-driven winds are important for understanding the physical mechanism and properties of mass loss on the asymptotic giant branch. These models are becoming increasingly realistic with more detailed physics included, but also more computationally demanding. The purpose of this study is to clarify to what extent the applied numerical approach affects resulting physical structures of modelled winds, and to discuss resulting changes. Following the previously developed radiation hydrodynamic model -- which includes descriptions for time-dependent dust formation and gas-dust drift -- and using its physical assumptions and parameters, numerical improvements are introduced. Impacts of the so-called adaptive grid equation and advection schemes are assessed from models calculated with different numerical setups. Results show that wind models are strongly influenced by numerical imprecision, displaying differences in calculated physical properties of up to one hundred per cent. Using a non-adaptive grid, models become periodic (in multiples of stellar pulsation periods), instead of irregular, as obtained previously. Furthermore, the numerical improvements reveal changes in physical structures. The influence of gas-dust drift is confirmed to be highly important, in particular for the dust component. Gas and dust are less tightly coupled than previously, and drastically larger amounts of dust form assuming drift.
\end{abstract}

\begin{keywords}
methods: numerical -- hydrodynamics -- radiative transfer -- stars: AGB and post-AGB -- stars: mass-loss -- stars: variables: general
\end{keywords}

\section{Introduction}\label{sec:introduction}
Dust-driven winds off of asymptotic giant branch (AGB) stars are complex physical phenomena that form in a highly non-linear interaction between a strong radiation field, a cool gas and conditions suitable for efficient dust formation. It is the radiation pressure on dust grains that with the support of levitated regions of an enclosed pulsating star is believed to form and drive more massive winds. Starting with \citet{Bo:88} a variety of time-dependent wind models, of increasing complexity, have been created in order to understand AGB star winds physically. An overview of properties and capabilities of recent wind models is given by, e.g., \citet[and references therein]{Ho:05}. \citet{Wi:00} reviews mass loss from cool stars in more general terms.

An important feature of self-consistent wind models, as presented here, is an adaptive grid equation. This equation distributes a fixed number of gridpoints across the model domain according to those physical properties that are required to be spatially resolved \citep[henceforth {\rDD}]{DoDr:87}. The intent to use an adaptive grid is sound because shocks form in radially steep density gradients of long period variables. Shocks that assist in providing appropriate conditions for dust formation and hence wind acceleration. It has earlier been found that a vast majority of all wind models using such a grid equation have formed irregular outflow patterns; for examples see: \citet[fig.~1]{HoDo:97}, \citet[henceforth {\rSaHoa}, fig.~2]{SaHo:03}, \citet[fig.~2]{HoGaArJo:03}, \citet[henceforth {\rSaHob}, fig.~5]{SaHo:03b}, and \citet[figs.~5--7]{MaHoHe:07}. Such irregularities could be understood as a result of physically (mildly) chaotic interactions, but they might also be a consequence of insufficient numerical accuracy.

Few studies address the use of the adaptive grid equation. To mention two \citet{Ku:94} compares results between models using first and second order accurate advection schemes; concluding that a second order advection is important. \citet[henceforth {\rDPSG}]{DoPiSt.:06} introduce both an improved discretization scheme and an improved advection scheme. In addition to the advection and discretization schemes, other ``numerical properties'' also have the potential of affecting the physical outcome of models. Such examples are how the adaptive grid equation is used, and the adopted grid resolution. There is no thorough study of wind models available in the literature, where the influence on results due to the numerical approach is explored. For models using the adaptive grid equation there is no such study available at all. This is an important issue of concern, since a model interpretation is uncertain when it is unclear to what degree the outcome is affected by numerical inaccuracy.

This paper is based on the work presented in the previous papers in the series, i.e., {\rSaHoa}, {\rSaHob}, and \citet[henceforth {\rSaHoc}]{SaHo:04}; a comprehensive summary is given in \citet{CSa:03}. The purpose of this paper is twofold. Firstly to fill the gap of a numerical study by exploring and determining a modelling approach by which numerical effects on results are separated and minimized. All physical parameters are thus fixed to previously used values. Secondly, the improved numerical approach leads to significant changes in the outcome, which are new and well worth addressing. For instance, when the adaptive grid does not track shocks outflow structures turn out to be periodic, instead of irregular as before. Assuming drift the wind also becomes more tenuous, simultaneously drastically larger amounts of dust are formed. Issues of the numerical method and physical results are treated together in order to emphasise how strongly they depend on each other. Results should, finally, also be valuable when improving related studies working with mass loss rates and yields of dust from AGB stars further, such as, e.g., \citet{FeGa:06}.

Numerical modifications and improvements are introduced, after first summarising the physical and numerical setup of the current AGB star wind model, in Sect.~\ref{sec:numerical}. The modelling procedure and model setup are then described in Sect.~\ref{sec:moprores}, where results and code consistency checks are also presented. The improved numerics in several cases gives rise to drastic changes to the physical structure. These changes and the revised outcome are discussed, after differences of the numerical modifications are treated, in Sect.~\ref{sec:discussion}. The paper is closed with conclusions in Sect.~\ref{sec:conclusions}.

\section{Model features and numerical improvements}\label{sec:numerical}
Following the three previous papers in this series, a distinction is made between three interacting physical components in the wind; gas, dust, and a radiation field. The system is described by coupled conservation equations, which account for exchange of mass, energy, and momentum between all three components -- forming the RHD3-system of equations (Radiation Hydrodynamics Dust and Drift). A thorough description of the physical system, the gas-dust interaction, and the numerical method was given in {\rSaHoa} (and references therein; where also a constant gas opacity was used), while {\rSaHob} added effects of stellar pulsations and grey gas opacities. {\rSaHoc}, finally, introduced effects of gas-dust drift in the dust formation process. The work presented here is based on this, most recent, physical formulation. All physical assumptions and parameters, as well as the general aspects of the numerical method, are unchanged with respect to these three papers, in order to focus on effects due to the numerical modifications studied here. In comparison to other AGB star wind models, these are based on -- and have most factors in common with -- the models presented by \citet{HoJoLoAr:98}.

\begin{table}
\caption{Glossary of used abbreviations and symbols}
\label{tab:glossary}
\begin{tabular}{lll}
\multicolumn{3}{l}{Abbreviations:}\\\hline\hline
term & \multicolumn{2}{l}{description}\\\hline
AGB  & \multicolumn{2}{l}{asymptotic giant branch}\\
RHD  & \multicolumn{2}{l}{radiation hydrodynamics}\\
RHDD & \multicolumn{2}{l}{radiation hydrodynamics \& time dependent dust formation}\\
RHD3 & \multicolumn{2}{l}{radiation hydrodynamics \& time dependent dust formation}\\
     & \multicolumn{2}{l}{-- including drift}\\
PC   & \multicolumn{2}{l}{position coupling (non-drift)}\\[1.0ex]
\multicolumn{3}{l}{\hspace*{0.3cm}Advection schemes; cf.\ Sect.~\ref{sec:numadv}}\\
\multicolumn{1}{r}{vL}   & \multicolumn{2}{l}{van Leer (second order)}\\
\multicolumn{1}{r}{VWvL} & \multicolumn{2}{l}{volume weighted van Leer (second order)}\\
\multicolumn{1}{r}{PPM}  & \multicolumn{2}{l}{piecewise parabolic method (third order)}\\[1.0ex]
\multicolumn{3}{l}{\hspace*{0.3cm}Grid types -- use of the adaptive grid equation}\\
\multicolumn{1}{r}{A}   & \multicolumn{2}{l}{adaptive and logarithmic distribution}\\
\multicolumn{1}{r}{L}   & \multicolumn{2}{l}{non-adaptive and logarithmic distribution}\\
\multicolumn{1}{r}{U}   & \multicolumn{2}{l}{non-adaptive and uniform distribution}\\[2.0ex]
\multicolumn{3}{l}{Symbols and properties:}\\\hline\hline
symbol & unit & description\\\hline
$r$    & cm                          & radius\\
$\rho$ & $\text{g}\,\text{cm}^{-3}$  & gas density\\
$e$    & $\text{erg}\,\text{g}^{-1}$  & specific internal energy of the gas\\
\teg   & K                           & gas temperature\\
$\kappa_{\text{g}}$ & $\text{cm}^2\text{g}^{-1}$ & gas opacity\\
$u$    & $\text{cm}\,\text{s}^{-1}$  & gas velocity\\
$J_\star$ & $\text{s}^{-1}\text{cm}^{-3}$ & net grain formation rate per volume\\
\taui  & $\text{s}^{-1}$             & net grain growth rate\\
\rhod  & $\text{g}\,\text{cm}^{-3}$  & dust density\\
\vdri  & $\text{cm}\,\text{s}^{-1}$  & drift velocity (size averaged)\\
$K_j$   & $\text{cm}^{-3}$            & moments of the grain size\\
        &                             & distribution; $0\le j\le3$\\
\fcond  &                             & degree of condensation\\
\drhog  &                             & dust-to-gas density ratio\\
$\drad$ & cm                          & mean grain radius\\
$R$     &                             & adaptive grid equation resolution function\\
$w$     &                             & adaptive grid equation weights for $R$\\[1.5ex]
\multicolumn{3}{l}{\hspace*{0.3cm}Model input parameters}\\
$L_\star$ & $\text{erg}\,\text{s}^{-1}$ & stellar luminosity\\
$M_\star$ & g                         & stellar mass\\
\Rs     & cm                          & stellar radius\\
\teff   & K                           & effective temperature\\
\CtoO   &                             & elemental abundance of C relative to O\\
$P$     & s                           & stellar pulsations: piston period\\
\deltaup & $\text{cm}\,\text{s}^{-1}$ & stellar pulsations: piston amplitude\\
\rint   & cm                          & radial location of the inner boundary\\
\rext   & cm                          & radial location of the outer boundary\\
\ngrid  &                             & number of gridpoints\\[1.5ex]
\multicolumn{3}{l}{\hspace*{0.3cm}Properties only calculated at the outer boundary}\\
$\dot{M}$ & \mdotu                    & mass loss rate\\
\uinf   & $\text{cm}\,\text{s}^{-1}$  & terminal velocity (final wind velocity)\\
\rfluc  &                             & relative fluctuation amplitude\\
$\langle q\rangle$ &                  & temporal mean of quantity $q$\\
\sistd  &                             & standard deviation\\[1.0ex]\hline
\end{tabular}
\end{table}

Gas-dust drift is considered ubiquitous in this article. However, since no other known AGB star wind model can handle drift without serious limitations, more restrictive non-drift (position coupled [PC]) models are also treated. This approach allows a more precise study than before of differences, both due to the numerical approach, and between drift and PC models.

New to this paper are modified descriptions of the spatial discretization and advection schemes, and how the adaptive grid equation is used. These modifications are introduced in Sects.~\ref{sec:numdisc}--\ref{sec:numgrid}, after a summary of the numerical method used so far is presented in Sect.~\ref{sec:nummethod}. Important issues related to the size of the model domain and grid resolution are additionally addressed in Sect.~\ref{sec:numngp}. All symbols and abbreviations used in this paper at more than one location are for a quick reference summarised in Table~\ref{tab:glossary}.

\subsection{Numerical method -- up to now}\label{sec:nummethod}
For a detailed description of the numerical method used so far see {\rSaHoa}, sect.~3.1, and references therein. Those features that are modified in this article are summarised in the following.

An adaptive grid equation \citep{DoDr:87} distributes gridpoints on the model domain by resolving gradients of selected quantities. Hence, gridpoints move back and forth according to the definition of a so-called grid resolution function $R$ (cf.\@ ibid.,\ eq.~3). The discretized and dimensionless form of $R$ can be written,
\begin{eqnarray}
R_i=\sqrt{1+\sum_{j=1}^Mw_j\left[\frac{X_i}{F_{j,i}}\frac{f_{j,i+1}-f_{j,i}}{r_{i+1}-r_i}\right]^2},
\label{eq:r}
\end{eqnarray}
where $X_i$ is a spatial scale length at gridpoint $i$, $r$ is the radius, $F_{j,i}$ the scaling factor for the physical quantity $f_{j,i}$ to be resolved (totally there are $M$ such quantities for each gridpoint), and $w_j$ are grid weights normally set to $1.0$ for the resolved quantities in stellar wind models (these grid weights are not written out explicitly in the original expression). Like before $R$ is set to be determined by the thermal (internal) energy ($e$) and the gas density ($\rho$). Abrupt changes in the distribution of gridpoints are, moreover, controlled using two smoothing factors. The temporal smoothing factor is set to \mbox{$\tau_{\mathrm{g}}=10^2$\,s}, which is smaller than, e.g., dynamical or dust related time scales in the wind, allowing the grid to freely adapt to physical features. The spatial smoothing factor is set to \mbox{$\alpha=2$}. Typically the gas density drops by a factor of $1.0$--$1.5$\,dex across a stronger shock why at least 10--15 gridpoints are required to achieve a sufficient local refinement (with $\alpha=2$, cf.~{\rDD}, p.~182).

Effects of stellar pulsations on the atmosphere and wind are described using a sinusoidal, radially varying inner boundary (of period $P$; i.e., a piston), located at about $\rint=0.91\,\Rs$; above the region where the $\kappa$-mechanism supposedly originates. An inflow of mass through the inner boundary -- like used by, e.g., \citet[][sect.~4.2]{SiIcDo:01} and \citet[henceforth {\rW}, sect.~2.5.1]{Wo:06} -- is as before not permitted. The typical amount of mass in the modelled envelope is about $0.002M_{\sun}$. Since the model domain is rather quickly depleted of material due to the wind, long-term modelling, covering thousands of years, is (currently) not possible.

The system of equations is discretized in a volume-integrated conservation form on a staggered mesh. Advection of mass, energy, and momentum between grid cells is described using a second order accurate discretization according to \citet{vLe:77}. To keep the formulation fully implicit the temporal discretization, of all terms in all equations, is always first order accurate. The equations are solved implicitly using a Newton-Raphson algorithm where the Jacobian of the system is inverted by the Henyey method.

\subsection{Discretization schemes -- introducing weighted means}\label{sec:numdisc}
The equations of radiation hydrodynamics in spherical geometry are discretized on a staggered mesh, i.e., scalar quantities such as the density are centred within a computational cell. Vector quantities, such as the velocity and radius, are localised to cell boundaries. When combining scalar and vector type quantities one must adopt a discretization scheme, which has up to now always been an arithmetic average of variables at neighbouring gridpoints (regardless of geometry). E.g.,
\begin{eqnarray}
r\mph=\frac{1}{2}\left(r_i+r_{i+1}\right)\quad\mbox{and}\quad\rho_i=\frac{1}{2}\left(\rho\mph+\rho\mpm\right),
\label{eq:numadvd}
\end{eqnarray}
where $r_i$ is the radius on the cell boundary at gridpoint $i$, and $r\mph$ the averaged radius at the cell centre ($r_{i+1}<r_i$). Other quantities are treated analogously. As pointed out by, e.g., {\rDPSG} (see sect.~2.1) a more consistent discretization scheme for a spherical geometry is achieved if quantities instead are weighted over the volume $V$ of each grid cell. Thus, the relations in Eq.~(\ref{eq:numadvd}) are replaced with,
\begin{eqnarray}
&&r\mph=\sqrt[3]{\frac{1}{2}\left(r^3_i+r^3_{i+1}\right)}\:\:\mbox{and}\nonumber\\
&&\:\:\:\:\:\:\rho_i=\frac{1}{2V_i}\left(\rho\mph V\mph+\rho\mpm V\mpm\right).
\end{eqnarray}

This improved discretization scheme has been implemented in all terms, and is used with all models using the volume weighted van Leer advection scheme (see next subsection). However, unlike the choice of the advection scheme (or use of the adaptive grid; cf.\@ Sect.~\ref{sec:numgrid}), the choice of the discretization scheme has in tests carried out for this paper been found to have a negligible influence on the physical structure of wind models.

\subsection{Advection schemes -- increasing the accuracy}\label{sec:numadv}
The adopted advection, describing transport of matter, energy, and momentum between grid cells, has up to now been the spatially second order accurate (van Leer [vL]) scheme of \citet[eq.~67]{vLe:77}\footnote{The drift models in {\rSaHoa} use a spatially first order accurate advection scheme, referred to as Donor Cell.}. In order to study the influence on stellar winds of higher accuracy in the advection, both the volume weighted second order (VWvL) scheme of {\rDPSG} as well as the third order piecewise parabolic method of \citet[PPM, which is also volume weighted; in the version of \citealt{WiNo:86} -- henceforth {\rWN} -- sect.~V-C]{CoWo:84}\footnote{Which inclusion is a capital task due to the large number of analytical derivatives required in an implicit formulation.} have been implemented. A temporal discretization only first order accurate motivates this use of higher order spatial discretization in the advection scheme, to reach an overall better accuracy.

Using the (integrated) mass in the advection term discretization (see, e.g., {\rWN}, sect.~V-B), the system Jacobian \citep[cf., e.g.,][sect.~5.3]{Do:98} becomes block penta-, hexa-, and nona-diagonal for the vL, VWvL, and PPM schemes, respectively. For the PPM scheme variables at the neighbouring gridpoints $(i-4,i-3,\hdots,i,\hdots,i+4)$ enter the discrete equations. Compared to the vL scheme the time penalties for each Gaussian elimination of the Jacobian are 20\,per cent (VWvL) and 80\,per cent (PPM).

In difference to the gas component it does not make sense to evaluate an integrated mass for the dust, which is why drift models always must be advected in the dust component using a velocity. With a velocity instead of a mass in the advection term discretization the corresponding Jacobian is block hexa-, hepta- and nona-diagonal. Time penalties are 20\,per cent (vL), 40\,per cent (VWvL), and 80\,per cent (PPM), respectively; when compared to the mass-advected vL scheme. It should be noted, however, that since the time-dependent iteration history changes with different advection schemes it is difficult to say exactly how the total model computation time changes. That the VWvL and PPM schemes have been correctly implemented is seen in Sect.~\ref{sec:disadv}, where both give very similar structures, although PPM achieves slightly more detail.

Up to now it has always been necessary to run adaptive grid wind models using the mass advection formulation in the gas equations, or the same equations fail to converge properly (resulting in very small time steps). In this context {\rDPSG} (in a different application) successfully switch from advection by mass to advection by velocity, after introducing the volume weighted discretization (Sect.~\ref{sec:numdisc}) and VWvL advection scheme. The same procedure \emph{does not} work (in general) with wind models using an adaptive grid. However, with a non-adaptive grid (see next subsection) wind models can too switch to the velocity advection formulation. Comparisons between mass and velocity advected models show negligible differences, which is why the mass formulation is used to save time in the calculation of PC models.

\subsection{The use of an adaptive or non-adaptive grid equation}\label{sec:numgrid}
Narrow shocks form in the steep density gradients of a pulsating AGB star atmosphere. The use of adaptive mesh refinement or an adaptive grid equation is motivated in order to resolve the physical structure of such shocks. There are, however, several drawbacks with possibly major consequences for the physical structure of the modelled region when an adaptive grid equation is used. Such drawbacks have not been addressed in any detail before and are treated in the following subsection. An approach using a non-adaptive grid (equation) is thereafter discussed in Sect.~\ref{sec:numgridna}.

\subsubsection{Issues of concern with an adaptive grid equation}\label{sec:numgridad}
While certain regions are adequately resolved through an appropriate definition of the grid resolution function, unresolved regions may, on the contrary, be ``gridpoint exhausted''. In particular if the model domain covers a spatially large region with a large number of structures. That is, structures may fail to form in a region where they could have formed if the same region was better resolved. This is of particular concern in drift models where large variations form in dust quantities between regions resolved by the grid equation. There are two immediate solutions to this problem, adding more gridpoints or using a different resolution function, but these solutions do not work well. Choosing a different resolution function in order to resolve quantities of the dust is difficult since dust properties primarily vary on shorter time scales than those of the gas; as the dust component lacks a pressure component and reacts very quickly to changes in the radiation field. Resolving both gas and dust features would at times require most of the model domain to be resolved simultaneously, disapproving its use.

Adding significantly more gridpoints is in general not an option, as the relative difference in the amount of mass between grid cells may become too large (since most gridpoints will be resolving shocks further), introducing new numerical problems. However, models using a mildly increased(/decreased) number of gridpoints should reproduce average properties, if the numerical scheme is stable (cf.\ Sect.~\ref{sec:numngp}). A third option is to use smaller grid weights in the grid resolution function ($w_j$ in Eq.~\ref{eq:r}), cf.\@ Sect.~\ref{sec:disgrid}.

There is a critical issue with the grid equation that is entirely limited to drift models. In a grid cell with a small amount of mass (like in the dust component) there is the possibility that the amount of numerical diffusion due to advection becomes too small to prevent numerical errors from growing out of control (see, e.g., sect.~IV-C in {\rWN}, and sect.~3.2 in {\rSaHoa}). This is the case in regions where gridpoints are moving in both directions to resolve physical (or numerical) features elsewhere and the local relative advection velocity becomes close to zero. In order to remedy the problem some artificial mass diffusion can be added to the dust component. An illustration of this issue can be seen in the dust velocities shown in {\rSaHoa}, figs.~2b \& j at $r=15\,\Rs$ and $r=25\,\Rs$ (where $\Rs$ is the stellar radius). Incidentally such numerical artefacts always occur in front of shocks where there is very little dust (and few gridpoints). If and when the dust drifts ahead of the gas these numerical artefacts together with a locally decreased resolution can cause major problems to the overall physical structure of the wind, cf.\@ Sect.~\ref{sec:dispvar}. This numerical issue makes drift models difficult to model with an adaptive grid equation. With a non-adaptive grid, which does not track shocks, the advection velocity is non-zero in regions with dust -- with the possible exception of cool inner regions -- and this problem is greatly reduced (artificial dust mass diffusion is still required occasionally).

A wind model covers a large spatial domain, so far typically $0.91$--$25\,\Rs$, with plenty of structures in both the gas and dust, which can all be important to the overall structure of the accelerating wind. In particular drift models form more structures in the dust than PC models do (cf., e.g., sect.~5.2 in {\rSaHob}). A large number of structures, which cannot all be resolved simultaneously, and the difficult numerical conditions for modelling drift are both good motives for using a non-adaptive grid. Still, one strong argument to keep the grid equation (in a non-adaptive setup) is that it allows a simple radial displacement of the inner boundary to simulate pulsations (cf., e.g., {\rSaHob}, sect.~2.3).

\subsubsection{Introducing a non-adaptive grid equation}\label{sec:numgridna}
On top of a slightly moving grid due to a displaced inner boundary (see above) the basic distribution of gridpoints can (most easily) be made logarithmic or uniform (henceforth grid types L and U). A logarithmic grid has the advantage that narrow shocks in the stellar centre are better resolved than with a uniform grid, also with fewer gridpoints. A logarithmic grid is therefore preferred over a uniform; the latter is, however, used for comparison.

Two relevant properties of the adaptive grid equation in this context are the resolution function $R$ (Eq.~\ref{eq:r}) and the gridpoint concentration $n$ (see {\rDD}, eq.~2). The grid equation is non-adaptive -- and not tracking shocks -- if $R$ is set to $1$. $n$ is defined by,
\begin{eqnarray}
n_i=\frac{X_i}{r_{i}-r_{i+1}},\quad\mbox{and}\quad X_{\text{L},i}=r^x\mph,\quad X_{\text{U},i}=C,
\end{eqnarray}
where $0\!\le\!x\!\le\!1$. A logarithmic grid is achieved if the scale length is set to $X=X_{\mathrm{L}}$ ($x\!=\!1$). Similarly a uniform distribution results if the scale length instead is set to $X=X_{\mathrm{U}}$, where $C$ is a constant. If $X=X_\text{L}$ and $x\!=\!0.5$ the grid will be very nearly uniform and non-adaptive (also with $R\ne1$), since the nominator and denominator of $n$ differ by orders of magnitude; using typical values of $r$. This setting is used with models using a spatially uniform grid.

With a non-adaptive grid time steps will necessarily be shorter on average since the ``stationary'' temporal periods found in models using an adaptive grid vanish. It simultaneously turns out that the required number of iterations in each time step of the non-adaptive model on the average is lower. In total the computational time required to reach a specific evolved ``age'' of a model is slightly higher for a non-adaptive model compared to the adaptive counterpart. As is found in Sect.~\ref{sec:results} all presented models using a non-adaptive grid form outflows with a periodic character instead of irregular, as was found previously. It is thus not necessary anymore to evolve a model to reach a high age in order to get sensibly temporally averaged quantities.

\subsection{Size of model domain and grid resolution}\label{sec:numngp}
Models typically span a region from about \mbox{$\rint=0.91\,\Rs$} to \mbox{$\rext=25$--60$\,\Rs$}, covering the atmosphere and wind acceleration region. In earlier papers in the series, the outer boundary was fixed at about $\rext\!=\!25\,\Rs$. {\rW} in his study places the outer boundary at \mbox{$\rext\!=\!10\,\Rs$}. The argument behind these values is that the wind has reached its terminal velocity within these domains.

The default number of gridpoints is $\ngrid=500$ for both models using an adaptive and non-adaptive grid; this is also the same number used previously. An issue related to the grid resolution is the amount of added artificial tensor viscosity. This amount is defined using a length scale (cf., e.g., \rSaHoa, eqs.~13 \& 14),
\begin{eqnarray}
l_\text{relative}=rf,\quad\text{vs.}\quad l_\text{fixed}=r_0f,
\end{eqnarray}
where $f$ is a constant that defines the shock width and $r_0$ another constant. In order to always resolve shocks in the innermost region with at least one gridpoint (in models using a logarithmic non-adaptive grid, with $\ngrid=500$ \& $\rext=50\,\Rs$) $l_\text{relative}$ is used with $f=7.0\times10^{-3}$. In models using a uniform grid $l_\text{fixed}$ is used instead, with $r_0=5\,\Rs$. Differences in the physical structure due to the chosen value on the length scale are found to be negligible for $f\le2.0\times10^{-2}$ (using P13C16U6$_{\ngrid=700}$). With larger values the physical structure changes from periodic to irregular.

Finally, the influence of the adopted number of gridpoints on the physical structure has not been studied in any detail with these stellar wind models earlier. Ideally models using an increased number of gridpoints should become more alike, if the numerical method (and physical circumstances) permits. In addition it is useful to see how dependent the physical structure and temporally averaged quantities are on a decreasing number of gridpoints. An accurate spatial discretization (Sect.~\ref{sec:numdisc}) and advection scheme (Sect.~\ref{sec:numadv}) are of increased value with less dense grids. To find out one set of model parameters has been run using models with four different number of gridpoints, cf.\@ Sect.~\ref{sec:results}.

\section{Modelling procedure and results}\label{sec:moprores}
In this section the modelling procedure is, at first, reviewed in Sect.~\ref{sec:respro}, followed by a description of the chosen physical setup and selected model parameters in Sect.~\ref{sec:respar}. Results are then presented in Sect.~\ref{sec:results}. A discussion of models using a constant gas opacity is thereafter given in Sect.~\ref{sec:resconst}. The outcome of some tests for code validation is provided in Sect.~\ref{sec:resvalid}. 

\subsection{Modelling procedure}\label{sec:respro}
The modelling procedure is as follows. A wind model is started from a hydrostatic dust-free initial model, where the outer boundary is located at about $\rext=2\,\Rs$. All dust equations and terms are then switched on simultaneously. Dust starts to form in the outer cooler domain, initiating an outward motion of dust and gas. The expansion is followed by the grid to a pre-defined radius (of about $\rext\!=\!25$--$60\,\Rs$), where the outer boundary is fixed and subsequent outflow allowed. Thereafter the actual wind is modelled.

A wind model is evolved for a time interval of about \mbox{$50$--$200\,P$}. The modelled time interval cannot be much shorter than \mbox{50$\,P$} (when \mbox{$\rext\!=\!25\,\Rs$}), which is the typical time needed for transients in the expansion phase to leave the model domain through the outer boundary. Temporally averaged quantities should be measured only after this time; and larger radial domains consequently require longer modelled time intervals. Measuring averaged quantities extended time intervals are unimportant for models using a non-adaptive grid, since these in all cases studied here are found to develop periodic outflows. For such models precise averaged quantities can be calculated over an interval of a few piston periods. The situation is different with models using an adaptive grid, which for most computed cases develop an irregular structure, like before. Such structures require a longer time interval to sensibly measure averaged quantities.

Models using a Planck mean gas opacity (see below) all develop a `low mass' envelope containing about $10^{-3}\,M_{\sun}$. The time interval used to compute averaged quantities must be short enough that the model domain is not depleted of mass (also see sect.~4.2 in {\rSaHob}). The intervals studied in Sect.~\ref{sec:results} are selected at times before a significant fraction of the envelope is lost; in comparison to a model with periodic structures it is in general more difficult to define a suitable short time interval with irregular structures.

\subsection{Physics setup and selection of model parameters}\label{sec:respar}
All models in this paper are selected with the purpose of studying relative changes compared to previously calculated wind models. This approach allows a quantitative and qualitative estimate of the importance of the adopted numerical modifications.

Highly critical to properties such as the developed density structure is the gas opacity used. Models adopting Planck mean absorption coefficients of molecular data (Planck mean opacities/models) on the one hand result in much more realistic density structures, compared to models calculated with a constant gas opacity (constant-opacity models; see, e.g., sect.~2.2 in {\rSaHob}). On the other hand Planck mean models are less realistic than models which use frequency-dependent opacities \citep[for both the gas and dust,][]{HoGaArJo:03}. Although frequency-dependent opacities introduce another level of complexity to the problem, and are also very computationally demanding. In the current context Planck mean gas opacities are used to avoid these problems.

In comparison to {\rSaHoc}, where the influence of various levels of drift inclusion in the dust formation equations was studied, all drift models here use the complete drift-dependent dust formation description (marked with the suffix ``$-dvs$'' in ibid.). One limitation of this formulation is that dust formation in principle is grain size dependent through the drift velocity (cf.\@ Sect.~\ref{sec:dispimprove}). The computational requirements of a study using a grain size distribution are much larger than with a mean grain radius, which is why such an approach is out of scope of this paper. Still, no other known time-dependent AGB star wind model includes drift-dependent dust formation, using even a mean grain size. All remaining physical parameters of the gas and dust are unchanged.

\begin{table}
\caption{Model parameters, see Sect.~\ref{sec:respar} for further details. The name of the model is given in Col.~1. The following five columns specify: the stellar luminosity $L_\star$, effective temperature {\teff}, pulsation period $P$, pulsation amplitude \deltaup, and carbon-to-oxygen ratio \CtoO. The last column shows the fraction of the stellar mass contained in the radial domain of the initial model. The stellar mass $M_\star$ is set to $1.0\,M_{\sun}$ in all models. These sets of model parameters have been taken from table~2 in {\rSaHob}.}
\label{tab:resmodpar}
\begin{tabular}{lcccccc}\hline\hline\\[-1.8ex]
model &$L_\star$       &{\teff}     &\pP         & \deltaup &\CtoO & $\displaystyle\frac{M_\text{e}}{M_\star}$\\
      &$[L_{\sun}]$&$[\mbox{K}]$&$[\mbox{d}]$&[km/s]& & [\%]\\[1.0ex]\hline\\[-1.8ex]
P10C16U6  & $1.0\times10^4$ & 2790 & 525 & 6 & 1.6 & 0.16\\[0.5ex]
P10C18U4  & $1.0\times10^4$ & 2790 & 525 & 4 & 1.8 & 0.13\\
P10C18U6  & $1.0\times10^4$ & 2790 & 525 & 6 & 1.8 & 0.13\\[0.5ex]
P13C14U6  & $1.3\times10^4$ & 2700 & 650 & 6 & 1.4 & 0.19\\[0.5ex]
P13C16U4  & $1.3\times10^4$ & 2700 & 650 & 4 & 1.6 & 0.16\\
P13C16U6  & $1.3\times10^4$ & 2700 & 650 & 6 & 1.6 & 0.16\\[1.0ex]\hline
\end{tabular}
\end{table}

The physical equations solved are highly non-linear, as is the outcome. A larger number of models have accordingly been calculated to provide better statistics. In order to enable a straightforward comparison with earlier results, the same model parameters used in {\rSaHob} (and {\rSaHoc}) have also been used here; see Table~\ref{tab:resmodpar}. Models for which a wind did not form earlier (PC models with a terminal velocity $\uinf\!<\!10\,\kms$) are also left out here, since they are not expected to form a wind now either. One model, P13C16U6, is used here in a thorough study of the influence of the number of gridpoints.

\subsection{Results}\label{sec:results}
In order to assess the importance of the numerical modifications, each set of model parameters has been used in a number of different setups. Both drift and PC models have been calculated for all sets. All models have additionally been calculated using at least two out of three advection schemes -- marked with either vL, VWvL, or PPM (see Sect.~\ref{sec:numadv}). Models using an adaptive grid (grid type A) have been recomputed for PC models using vL and VWvL advection. Both to see what the influence of the advection scheme is, and to see how new values of recomputed vL-models match previously calculated values -- as a consistency check (cf.\ Sect.~\ref{sec:resvalid}). Drift models have not been calculated at all here with an adaptive grid due to the numerical difficulties mentioned in Sect.~\ref{sec:numgridad}. Moreover, all models using a non-adaptive grid have been calculated using both a logarithmic (L) and a uniform (U) grid type. Since the emphasis has been to resolve shocks, only results of models using a logarithmic grid are presented; except for P13C16U6.

\begin{table*}
\caption{Temporally averaged quantities at the outer boundary; see Sect.~\ref{sec:results}. Several models with different numerical setups are presented for each set of modelled parameters. From the left the first five columns specify input properties: model name, PC or drift (p/d), advection scheme (A), grid type (g), and radius at the outer boundary ($r_\text{ext}$). Remaining columns characterise the outflow in seven properties: the mass loss rate {\mmdot}, terminal velocity {\muinf}, degree of condensation {\mfcond}, dust/gas density ratio {\mdrhog}, dust radius {\mdrad}, drift velocity {\mvdri} (only for drift models), and type of outflow (type) for the gas:dust. Advection schemes (vL, VWvL and PPM) are described in Sect.~\ref{sec:numadv}, and grid types (A, L and U) likewise in Sect.~\ref{sec:numgrid}. The types of outflow structures that form are classified as: s, stationary; i, irregular; $l$\,p, periodic; and $l$\,q quasi-periodic. $l$ ($\in\mathbb{N}$) indicates the (multi-)periodicity of the gas/dust outflow in the unit of the piston period $P$. Models using an adaptive grid all show the same type of structure in both the gas and dust. In addition a relative fluctuation amplitude $\rfluc\,(=\!\sistd/q$; {\sistd} is the standard deviation) is specified for each quantity $q$ (cf.~Sect.~\ref{sec:results}). The most accurately calculated drift and PC (reference) models are marked with \textsc{ref.} in Col.~1. Values shown in boldface/underline indicate a significant difference ($\ge10$\,per cent) from the corresponding drift/PC reference model value.}
\label{tab:restaq}
\begin{tabular}{l@{\quad\quad}%
                c@{\quad}r@{\ \ }c@{}cr%
                @{\ \ }r@{}rr%
                @{\ \ }r@{}rr%
                @{\ \ }r@{\ \ }rr%
                @{\ \ }r@{\ \ }rr%
                @{\ \ }r@{\ \ }rr%
                @{\ \ }r@{\ }rr}%
              \hline\hline\\[-1.8ex]
   \multicolumn{1}{l}{model} & p/d & \multicolumn{1}{c}{A} & $g$ & {\rext} &&
          \multicolumn{2}{c}{$10^6$\,\mmdot} &&
          \multicolumn{2}{c}{\muinf}  &&
          \multicolumn{2}{c}{\mfcond} &&
          \multicolumn{2}{c}{\mdrhog} &&
          \multicolumn{2}{c}{$10^2$\mdrad}  &&
          \multicolumn{2}{c}{\mvdri}  & type\\
      &&&& [\Rs] &&\multicolumn{2}{c}{$[\mdotu]$} &&
          \multicolumn{2}{c}{$[\kms]$}    &&
          \multicolumn{2}{c}{$[\%]$}      &&
          \multicolumn{2}{c}{$[10^{-4}]$} &&
          \multicolumn{2}{c}{$[\mu\text{m}]$}      &&
          \multicolumn{2}{c}{$[\kms]$}\\
      &&&&&&&$10^{3}$\rfluc &&
          &$10^{3}$\rfluc &&
          &$10^{3}$\rfluc &&
          &$10^{3}$\rfluc &&
          &$10^{3}$\rfluc &&
          &$10^{3}$\rfluc\\[1.0ex]\hline\\[-1.0ex]
P10C16U6 & PC &   vL & A & 25 &&\bfs{2.9}  & {\it 480}  & & \bfs{13}  & {\it 85}  & & \bfs{25}  & {\it 160} & & \bfs{8.6} & {\it 160} && {\bf3.0}  & {\it 150} &&&& \bfs{i}\\[0.15ex]
         &    & VWvL & A & 25 &&\bfs{2.5}  & {\it 760}  & & \bfs{14}  & {\it 93}  & & \bfs{28}  & {\it 170} & & \bfs{9.6} & {\it 180} && \bfs{3.2} & {\it 180} &&&& \bfs{i}\\[0.15ex]
         &    &   vL & L & 32 &&\bfs{2.45} & {\it 7.1}  & &  {\bf10.4} & {\it1.2} & & {\bf18.2} & {\it 6.6} & & {\bf6.16} & {\it6.1}  && {\bf3.02} & {\it 5.0} &&&& {\bf s}:\bfs{s}\\[0.15ex]
\mref    &    & VWvL & L & 32 && {\bf1.69} & {\it 21}   & &  {\bf10.9} & {\it2.5} & & {\bf18.7} & {\it 2.7} & & {\bf6.42} & {\it2.7}  && {\bf2.79} & {\it 1.8} &&&& {\bf s}:{\bf q}\\[1.0ex]
      & drift      &   vL & L & 40 &&     1.31 & {\it 440}   & &      14.5  & {\it 48} & & {\bf27.7} & {\it 990} & & {\bf 11.4} & {\it1200} && {\bf3.30} & {\it460} && {\bf8.82} & {\it 460}  & {\bf 5q:5q}\\[0.15ex]
\mref    &    & VWvL & L & 40 &&     1.24 & {\it 270}   & &      15.8  & {\it 16} & &     38.7  & {\it 760} & &      16.4  & {\it1000} &&     4.86  & {\it140} &&     4.99  & {\it 640}  & 2p:2p\\[2.0ex]\cline{1-5}\\
P10C18U4 & PC &   vL & A & 25 &&     1.2 & {\it 470} & &  {\bf15} & {\it 41} & & {\bf17} & {\it 140}& & {\bf7.7} & {\it 140}   && {\bf1.7} & {\it 120} &&&&  \bfs{i}\\[0.15ex]
         &    & VWvL & A & 25 &&     1.2 & {\it 480} & &  {\bf14} & {\it 59} & & {\bf17} & {\it 190}& & {\bf7.6} & {\it 200}   && {\bf1.7} & {\it 110} &&&&  \bfs{i}\\[0.15ex]
         &    & VWvL & L & 40 &&     1.08 & {\it 12}  & &  {\bf14.4} & {\it1.4} & & {\bf16.1} & {\it 8.9} & & {\bf7.36} & {\it 8.7}   && {\bf1.69} & {\it 4.4} &&&&  \bfs{s}:\bfs{s}\\[0.15ex]
\mref    &    &  PPM & L & 40 &&     1.08 & {\it 40}  & &  {\bf14.6} & {\it4.3} & & {\bf16.3} & {\it 12}  & & {\bf7.44} & {\it 12}    && {\bf1.69} & {\it 11}  &&&&       1p:1p\\[1.0ex]
\mref &drift   & VWvL & L & 50 &&     1.10 & {\it 21}  & &      20.3  & {\it1.3} & &     52.1  & {\it 68}  & &      21.2 & {\it 140}   &&    3.30  & {\it9.7}  &&      4.99  & {\it 44} & 1p:1p\\[2.0ex]\cline{1-5}\\
P10C18U6 & PC &   vL & A & 25 &&     2.5 & {\it 400} & & \bfs{17} & {\it 42} & &\bfs{23} & {\it 120} & & \bfs{10} & {\it 130} && \bfs{2.1}& {\it 100} &&&&  \bfs{i}\\[0.15ex]
         &    & VWvL & A & 25 &&     2.4 & {\it 710} & & \bfs{18} & {\it 61} & &\bfs{27} & {\it 220} & & \bfs{12} & {\it 230} && \bfs{1.9}& {\it 34}  &&&&  \bfs{i}\\[0.15ex]
         &    &   vL & L & 40 &&\bfs{3.17}& {\it 18}  & & \bfs{16.8} & {\it1.6} & &\bfs{23.9} & {\it 4.8} & & \bfs{10.8} & {\it 4.8} && \bfs{2.45}& {\it 3.2} &&&&  \bfs{s}:\bfs{s}\\[0.15ex]
         &    & VWvL & L & 40 &&     2.41 & {\it 460} & &  {\bf20.0} & {\it27}  & &\bfs{42.6} & {\it 190} & & \bfs{19.4} & {\it 190} && {\bf3.02} & {\it 120} &&&&  {\bf2p}:{\bf2p}\\[0.15ex]
\mref    &    &  PPM & L & 40 &&     2.34 & {\it 550} & &  {\bf20.9} & {\it34}  & & {\bf50.1} & {\it 230} & &  {\bf22.9} & {\it 230} && {\bf3.26} & {\it 150} &&&&  {\bf2p}:{\bf2p}\\[1.0ex]
      & drift   &   vL & L & 50 && {\bf2.88}& {\it 500} & &      24.8  & {\it 22} & & {\bf43.7} & {\it 750} & &  {\bf23.9} & {\it 1200}&& 4.42 & {\it 46}  &&{\bf-0.286} & {\it 18000}   & {\bf2p}:{\bf 2p}\\[0.15ex]
\mref  &       & VWvL & L & 50 &&     2.28 & {\it 54}  & &      23.5  & {\it2.8} & &     65.5  & {\it 95}  & &      27.2  & {\it 260} && 4.38 & {\it 16}  &&      4.40  & {\it 150} & 1p:1p\\[2.0ex]\cline{1-5}\\
P13C14U6 & PC &   vL & A & 32 &&\bfs{6.0} & {\it 280}& & \und{11} & {\it 45} & & \bfs{22}& {\it 50}  & &\bfs{5.1} & {\it 49}  && {\bf5.0} & {\it 38} &&&&  \bfs{i}\\[0.15ex]
         &    & VWvL & A & 31 &&\bfs{4.2} & {\it 95} & & {\bf7.5} & {\it 15} & &  {\bf18}& {\it 53}  & & {\bf4.1} & {\it 54}  && \bfs{5.3}& {\it 26} &&&&  \bfs{i}\\[0.15ex]
         &    &   vL & L & 40 &&\bfs{3.86} & {\it 14} & & {\bf7.40} & {\it 3.3} & & {\bf16.1} & {\it 11}   & & {\bf3.65} & {\it 11}   && {\bf4.84} & {\it5.5} &&&&  s:{\bf s}\\[0.15ex]
         &    & VWvL & L & 40 && {\bf2.83} & {\it 6.5}& & {\bf7.78} & {\it 0.76}& & {\bf17.2} & {\it 3.2}  & & {\bf3.92} & {\it 3.2}  && {\bf4.73} & {\it0.087}&&&&  s:{\bf s}\\[0.15ex]
\mref    &    &  PPM & L & 51 && {\bf2.66} & {\it 2.0}& & {\bf7.88} & {\it 0.89}& & {\bf17.3} & {\it 1.8}  & & {\bf3.95} & {\it 1.8}  && {\bf4.72} & {\it0.54}&&&&  s:{\bf s}\\[1.0ex]
         & drift  &   vL & L & 55 && {\bf1.79} & {\it 190}& &      10.2 & {\it 16} & & {\bf32.0} & {\it 650} & & {\bf6.56} & {\it 860} && {\bf5.93} & {\it180} && {\bf6.51} & {\it 190} &  {\bf i}:{\bf i}\\[0.15ex]
\mref    &        & VWvL & L & 55 &&     2.21  & {\it 7.1}& &      10.2  & {\it 15}& &     42.8  & {\it 110} & &     8.28  & {\it 200} &&     7.78  & {\it9.5} && 4.09  & {\it 90}& s:2p\\[2.0ex]\cline{1-5}\\
P13C16U4 & PC &   vL & A & 30 &&\bfs{3.1} & {\it 42} & & \bfs{13} & {\it 7.7}& &\bfs{13} & {\it 27} & &\bfs{4.4} & {\it 34} && \bfs{2.7}& {\it 17} &&&&  \bfs{s}\\[0.15ex]
         &    & VWvL & A & 30 &&\bfs{3.1} & {\it 58} & & \bfs{13} & {\it 5.4}& &\bfs{15} & {\it 21} & &\bfs{5.0} & {\it 22} && {\bf2.9} & {\it 12} &&&&  \bfs{s}\\[0.15ex]
         &    & VWvL & L & 51 && {\bf3.57} &{\it 6.1} & &  {\bf14.6} & {\it0.69}& & {\bf19.1} & {\it 2.9}& & {\bf6.54} & {\it 2.9}&& {\bf3.11} & {\it 1.4}&&&&  s:\bfs{s}\\[0.15ex]
\mref    &    &  PPM & L & 51 && {\bf3.69} &{\it 9.9} & &  {\bf14.7} & {\it 1.1}& & {\bf19.8} & {\it 5.1}& & {\bf6.78} & {\it 5.4}&& {\bf3.17} & {\it 3.4}&&&&  s:1p\\[1.0ex]
      & drift &   vL & L & 56 && {\bf2.47} & {\it240}& &    18.6  & {\it 10} & &    53.0 & {\it310} & &     16.6 & {\it 560}&&   5.07   & {\it 45} &&     3.34 & {\it 320}  & {\bf2p}:{\bf2p}\\[0.15ex]
\mref    &    & VWvL & L & 56 &&     2.78  &{\it 9.5}& &    17.8  & {\it0.43}& &    54.5 & {\it 6.8}& &     16.8 & {\it 15} &&   4.99   & {\it1.4} &&     3.47 & {\it 43}   & s:1p\\[2.0ex]\cline{1-5}\\
P13C16U6 & PC &   vL & A & 30 && {\bf4.9} &{\it 160} & & {\bf15}  & {\it 200}& & \bfs{21}& {\it 48} & &\bfs{7.0} & {\it 500} && \bfs{3.1}& {\it 35} &&&&  \bfs{i}\\[0.15ex]
         &    & VWvL & A & 30 && {\bf5.2} & {\it250} & & {\bf15}  & {\it 23} & & \bfs{21}& {\it 86} & &\bfs{7.1} & {\it 86}  && \bfs{3.1}& {\it 61} &&&&  \bfs{i}\\[0.15ex]
         &    & VWvL & L & 51 &&     4.84  & {\it 2.9}& & {\bf15.9}  & {\it1.3} & &  {\bf24.5}& {\it 2.6} & & {\bf8.39} & {\it 2.6} && {\bf3.56} & {\it 2.3}&&&&  s:{\bf s}\\[0.15ex]
\mref    &    &  PPM & L & 51 && {\bf4.97} & {\it 14} & & {\bf16.1}  & {\it1.4} & &  {\bf25.5}& {\it 2.4} & & {\bf8.75} & {\it 2.4} && {\bf3.65} & {\it 2.0}&&&&  s:{\bf s}\\[0.15ex]
         &    &  PPM & U & 30 &&     4.71  & {\it 103}& & {\bf15.6}  & {\it 9.3}& & {\bf24.4} & {\it 28}  & & {\bf8.35} & {\it 27}  && {\bf3.45} & {\it 74} &&&&  \bfs{1p}:\underline{1p}\\[1.0ex]
      & drift   &   vL & L & 50 &&     4.72 & {\it400} & &    20.6   & {\it 20} & &  {\bf55.1}& {\it 530} & & {\bf22.4} & {\it 880} && 5.88 & {\it 21} && {\bf2.60} & {\it 970} & {\bf2p}:{\bf2p}\\[0.15ex]
\mref &         & VWvL & L & 55 &&     4.40 & {\it 14} & &    19.9   & {\it0.66}& &      62.5 & {\it 16}  & &     19.4  & {\it 42}  && 6.11 & {\it 4.5}&&     3.40  & {\it 110} & s:1p\\[0.15ex]
      &         & VWvL & U & 30 &&     4.01 & {\it 90} & &    19.7   & {\it 6.5}& &  {\bf56.2}& {\it 300} & &      19.3 & {\it 600} && 5.89 & {\it  25}&&     3.29  & {\it 220} & {\bf1p}:1p\\[2.0ex]\hline\\[-1.0ex]
\end{tabular}
\end{table*}

\begin{table*}
\caption{Properties temporally averaged at the outer boundary for model P13C16U6 (see Sect.~\ref{sec:results}) using different numbers of gridpoints. From the left columns specify input properties: PC or drift type, number of gridpoints {\ngrid}, advection scheme (A), and grid type (g). Remaining columns specify seven properties: the mass loss rate {\mmdot}, terminal velocity {\muinf}, degree of condensation {\mfcond}, dust/gas density ratio {\mdrhog}, mean grain radius {\mdrad}, drift velocity {\mvdri}, and type of outflow (type). All models have been calculated with the outer boundary fixed at $\rext\!=\!50\,\Rs$. For further details see the caption of Table~\ref{tab:restaq}.}
\label{tab:restag}
\begin{tabular}{l@{\ \ }cr@{\ \ }cc@{\qquad\quad}%
    r@{\ }rc%
    r@{\ \ }rc%
    r@{\ }rc%
    r@{\ }rc%
    r@{\ }rc%
    r@{\ }rc%
    r}\hline\hline\\[-1.8ex]
   \multicolumn{1}{l}{drift/PC} & \ngrid & \multicolumn{1}{c}{A}& $g$ &&
          \multicolumn{2}{c}{$10^6$\,\mmdot} &&
          \multicolumn{2}{c}{\muinf}  &&
          \multicolumn{2}{c}{\mfcond} &&
          \multicolumn{2}{c}{\mdrhog} &&
          \multicolumn{2}{c}{$10^2$\,\mdrad} &&
          \multicolumn{2}{c}{\mvdri}  & type\\
   \multicolumn{1}{r}{model} &  &&&&\multicolumn{2}{c}{$[\mdotu]$} &&
          \multicolumn{2}{c}{$[\kms]$}    &&
          \multicolumn{2}{c}{$[\%]$}      &&
          \multicolumn{2}{c}{$[10^{-4}]$} &&
          \multicolumn{2}{c}{$[\mu\text{m}]$}  &&
          \multicolumn{2}{c}{$[\kms]$}\\
      &&&&&&$10^3\rfluc$ &&
          &$10^3\rfluc$ &&
          &$10^3\rfluc$ &&
          &$10^3\rfluc$ &&
          &$10^3\rfluc$ &&
          &$10^3\rfluc$\\[1.0ex]\hline\\[-1.0ex]
PC    & 700 & vL   & A && {\bf5.3}  & {\it 200}   & & {\bf15}    & {\it 25}    & & \bfs{22}  & {\it 33}    & & \bfs{7.5} & {\it 32}    && {\bf3.3}  & {\it 28} &&&& \bfs{i}\\[0.15ex]
          & & VWvL & A &&\bfs{6.0}  & {\it 260}   & & {\bf15}    & {\it 22}    & & \bfs{22}  & {\it 75}    & & \bfs{7.6} & {\it 75}    && {\bf3.4}  & {\it 52} &&&& \bfs{i}\\[0.15ex]
          & & vL   & L &&\bfs{5.49} & {\it 7.4}   & & {\bf15.3}  & {\it 25}    & & \bfs{22.6}& {\it 5.8}   & & \bfs{7.72}& {\it 5.8}   && {\bf3.51} & {\it 2.7}&&&& s:{\bf s}\\[0.15ex]
          & & VWvL & L && {\bf5.06} & {\it 20}    & & {\bf16.1}  & {\it 2.0}   & &  {\bf25.4}& {\it 7.1}   & &  {\bf8.70}& {\it 7.1}   && {\bf3.66} & {\it 7.4}&&&& s:{\bf s}\\[0.15ex]
          & & PPM  & L &&     4.99  & {\it 50}    & & {\bf16.2}  & {\it 4.3}   & &  {\bf25.6}& {\it 8.5}   & &  {\bf8.76}& {\it 8.7}    && {\bf3.65} & {\it 15} &&&& \bfs{1p}:\und{1p}\\[1.0ex]
drift &     &   vL & L &&     4.57  & {\it 630}   & &     21.0   & {\it 27}    & &  {\bf45.2}& {\it 810}   & &  {\bf23.8}& {\it 1200}  && 5.60      & {\it 85} && {\bf2.56} & {\it1300} & {\bf2p}:{\bf2p}\\[0.15ex]
      &     & VWvL & L &&     4.54  & {\it 76}    & &     19.8   & {\it 4.9}   & &      61.1 & {\it 130}   & &      19.4 & {\it 320}   && 6.09      & {\it 18} &&     3.26  & {\it 140} & {\bf 1p}:1p\\[1.0ex]\cline{1-5}\\
PC    & 500 &   vL & A && {\bf4.9} & {\it 160}   & & {\bf15}  & {\it 20}    & & \bfs{21}& {\it 48}    & & \bfs{7.0}& {\it 500}   && \bfs{3.1} & {\it 36} &&&&  \bfs{i}\\[0.15ex]
          & & VWvL & A && {\bf5.2} & {\it 250}   & & {\bf15}  & {\it 23}    & & \bfs{21}& {\it 86}    & & \bfs{7.1}& {\it 860}   && \bfs{3.1} & {\it 61} &&&&  \bfs{i}\\[0.01ex]\\[-2ex]
          & &   vL & L &&     4.79  & {\it 10}    & & {\bf15.9}  & {\it 0.46}  & &  {\bf24.4}& {\it 2.9}   & & {\bf8.37} & {\it 2.9}   && {\bf3.55} & {\it 2.6} &&&& s:{\bf s}\\[0.15ex]
          & & VWvL & L &&     4.84  & {\it 2.9}   & & {\bf15.9}  & {\it 1.3}   & &  {\bf24.5}& {\it 2.6}   & & {\bf8.39} & {\it 2.6}   && {\bf3.56} & {\it 2.3} &&&& s:{\bf s}\\[0.15ex]
\mref   &   &  PPM & L && {\bf4.97} & {\it 14}    & & {\bf16.1}  & {\it 1.4}   & &  {\bf25.5}& {\it 2.4}   & & {\bf8.75} & {\it 2.4}   && {\bf3.65} & {\it 2.0} &&&& s:{\bf s}\\[2.0ex]
drift &     &   vL & L &&     4.72  & {\it 400}   & &     20.6   & {\it 20}    & &  {\bf55.1}& {\it 530}   & & {\bf 22.4} & {\it 880}  &&     5.88  & {\it 21}  && {\bf2.60} & {\it 970} & {\bf2p}:{\bf2p}\\[0.15ex]
\mref   &   & VWvL & L &&     4.40  & {\it 14}    & &     19.9   & {\it 0.66}  & &     62.5  & {\it 16}    & &      19.4  & {\it 42}   &&     6.11  & {\it 4.5} &&     3.40  & {\it 110} & s:1p\\[1.0ex]\cline{1-5}\\
PC    & 300 &   vL & A &&\bfs{6.0}  & {\it 63}    & & {\bf15}  & {\it 2.1}   & &\bfs{21} & {\it 18}    & &\bfs{7.1} & {\it 18}    && {\bf3.3} & {\it 8.3} &&&&  \bfs{i}\\[0.15ex]
          & & VWvL & A &&\bfs{6.1}  & {\it 65}    & & {\bf15}  & {\it 6.3}   & &\bfs{21} & {\it 39}    & &\bfs{7.0} & {\it 39}    && {\bf3.3} & {\it 21}  &&&&  \bfs{i}\\[0.01ex]\\[-2ex]
          & &   vL & L &&\bfs{6.15} & {\it 53}    & & {\bf14.6}  & {\it 5.7}   & &\bfs{20.1} & {\it 3.4}   & &\bfs{6.84} & {\it 34}    && {\bf3.40} & {\it 17}  &&&&  \bfs{9p}:\bfs{9p}\\[0.15ex]
          & & VWvL & L &&     4.75  & {\it 4.3}   & & {\bf15.7}  & {\it 0.83}  & & {\bf23.8} & {\it 1.6}   & & {\bf8.14} & {\it 1.6}   && {\bf3.51} & {\it 1.0} &&&&  s:{\bf s}\\[0.15ex]
          & &  PPM & L &&     4.76  & {\it 6.1}   & & {\bf15.9}  & {\it 1.1}   & & {\bf24.4} & {\it 2.3}   & & {\bf8.37} & {\it 2.2}   && {\bf3.56} & {\it 2.5} &&&&  {\bf1p}:\und{1p}\\[2.0ex]
drift &     &   vL & L && {\bf5.06} & {\it 100}   & &      18.9  & {\it 6.5}   & &     58.0  & {\it 200}   & &      18.9  & {\it 420}   && 6.11 & {\it 12}  && {\bf4.26} & {\it 230} & {\bf2p}:{\bf2p}\\[0.15ex]
      &     & VWvL & L &&     4.21  & {\it 120}   & &      20.0  & {\it 11}    & &     62.5  & {\it 140}   & &     20.2   & {\it 350}   && 6.02 & {\it 9.5} &&     3.55  & {\it 190} & {\bf2p}:{\bf2p}\\[1.0ex]\cline{1-5}\\
PC    & 100 &   vL & A && {\bf5.1} & {\it 30}    & &\bfs{13}  & {\it 4.0}   & &\bfs{16} & {\it 10}    & &\bfs{5.6} & {\it 10}    && \bfs{2.9} & {\it 6.0}   &&&&  s\\[0.15ex]
          & & VWvL & A && {\bf5.4} & {\it 20}    & & {\bf15}  & {\it 1.1}   & &\bfs{21} & {\it 2.2}   & &\bfs{7.3} & {\it 2.2}   && {\bf3.3} & {\it 3.5}   &&&&  s\\[0.01ex]\\[-2ex]
          & &   vL & L && {\bf4.96} & {\it 13}    & &\bfs{14.4}  & {\it 3.8}   & &\bfs{15.9} & {\it 7.1}   & &\bfs{5.35} & {\it 7.3}   && \bfs{2.58} & {\it 4.9}  &&&&  s:{\bf s}\\[0.15ex]
          & & VWvL & L &&\bfs{3.66} & {\it 3.3}   & &\bfs{13.8}  & {\it0.18}   & &\bfs{17.0} & {\it 0.13}  & &\bfs{5.83} & {\it 0.14}  && \bfs{2.93} & {\it 0.24} &&&&  s:{\bf s}\\[0.15ex]
          & &  PPM & L &&\bfs{4.16} & {\it 17}    & & {\bf14.7}  & {\it0.91}   & &\bfs{20.4} & {\it 2.5}   & &\bfs{6.97}  & {\it 2.3}   && {\bf3.34} & {\it 3.4} &&&&  s:{\bf s}\\[2.0ex]
drift &     &   vL & L &&     4.15  & {\it 150}   & &      20.1  & {\it 10}    & & {\bf45.2} & {\it 110}   & & {\bf13.5}  & {\it 210}   && {\bf4.88} & {\it 33} && {\bf8.66}& {\it 90} & s:{\bf s}\\[1.0ex]\hline\\[-1.0ex]
\end{tabular}
\end{table*}

The wind is characterised through temporally averaged properties, such as the mass loss rate {\mmdot} and terminal velocity {\muinf} for the gas. These two quantities together give the gas density $\langle\rho\rangle$. The dust is characterised through four properties: the degree of condensation {\mfcond}, dust-to-gas density ratio {\mdrhog}, mean grain radius {\mdrad}, and drift velocity {\mvdri} (only for drift models). The degree of condensation is defined as,
\begin{eqnarray}
\fcond=\frac{\rhod}{\rho_{\text{c}}^{\text{tot}}}\approx\frac{K_3}{K_3+n_\text{C}},
\end{eqnarray}
where $\rhod=m_1K_3$ is the dust density, and $K_3$ the third moment of the grain size distribution ($m_1$ is the dust grain monomer mass; cf., e.g., {\rSaHoa}, sect.~2.1); and $\rho_{\text{c}}^{\text{tot}}$ the total density of condensible matter (present in both the gas and dust phases); $n_{\text{c}}$ the total number density of condensible material in the gas phase. Providing a measure for the variability of the model structure, each outflow property ($q$) is attended by a relative fluctuation amplitude ($\rfluc\!=\!\sistd/q$; where {\sistd} is the [sample] standard deviation of the quantity $q$ in the measured time interval).

Results for models using the selected model parameters, different advection schemes, and grid types are presented in Table~\ref{tab:restaq}. In order to emphasise the increased accuracy of models using a non-adaptive grid, the corresponding values are written out with three significant digits. For each set of model parameters PC models are listed before drift models. Additionally, for each set the two drift and PC models calculated using the best available advection scheme and a non-adaptive grid are treated and marked as reference models (\textsc{ref.}). Other models in a set are marked in boldface if a property deviates by more than 10\,per cent (i.e., significantly) from the corresponding reference drift model value. Likewise, PC models are underlined if a property deviates by more than 10\,per cent from the respective PC reference model value.

In order to clarify differences all properties are in addition to Table~\ref{tab:restaq} also illustrated in Fig.~\ref{fig:disphys}; for absolute values (top panels) as well as values normalised to the respective value of each PC model (bottom panels). For each quantity ($q$) error bars in the figure show the property,
\begin{eqnarray}
\hat{e}=10^3\times\log_{10}\left[\hat{r}(q)-\max \hat{r}_\text{dr}(q)\right],
\end{eqnarray}
where $\hat{r}_\text{dr}(q)$ indicates a normalisation to the respective drift model value. Due to significant variations of the relative fluctuation amplitude for separate properties in different models this representation allows a visually better comparison.

The influence of the grid resolution is studied for model P13C16U6 using four different numbers of gridpoints: {\ngrid}=100, 300, 500, and 700 (the outer boundary is fixed at $\rext\!\simeq\!50\,\Rs$). This model was part of a detailed study in {\rSaHoc}. It is chosen for discerning differences arising from different resolutions, since it shows fairly small differences due to different advection schemes and grid types with {\ngrid}=500. Results are given in Table~\ref{tab:restag}. The number of gridpoints is printed in subscript following the model name when referring to these models.

\subsection{Comparing outcome with (grey) constant-opacity models}\label{sec:resconst}
Several constant-opacity models have also been calculated with all numerical modifications applied. As a contrast to the Planck mean models of this study a majority of the constant-opacity models develop an irregular structure, both with an adaptive and non-adaptive grid. Differences between outflow properties of the two models appear to be small. That these models form irregular structures, also when using a non-adaptive grid, is in good agreement with the latest results of {\rW}. He computes constant-opacity one-dimensional wind models using a code with adaptive mesh refinement.

Moreover, by calculating the model P13C16U6 with a pre-defined constant gas opacity -- such as $\kappa_{\text{g}}\!=\!4\times10^{-2}\,\text{g}^{-1}\text{cm}^2$ -- an envelope results that has a very similar amount of mass as the original model. Compared to the periodic structure of the original Planck mean model this constant-opacity model forms a wind with irregular structures. The conclusion is that constant-opacity models are more ``chaotic'' and less sensitive to numerical details than Planck mean models are; because of the fewer degrees of freedom, possibly in combination with very massive envelopes.

\subsection{Consistency checks -- code validation}\label{sec:resvalid}
Large modifications have been introduced to the code since its last application in {\rSaHoc}. It is, hence, important to compare new values with old to see how similar they are. New models have been calculated for all sets of PC model parameters using an adaptive grid; these are marked with vL-A in Table~\ref{tab:restaq} (Cols.~3 \& 4). The values should be compared with those values of the corresponding model in {\rSaHob} (see table~5 therein). For the P10-prefixed models the agreement is within 10\,per cent for all three models and quantities, except for the mass loss rate of P10C18U6, which is 25\,per cent larger here. The disagreement is bigger for the P13-prefixed models. P13C16U6, first, shows the same agreement as P10C18U6 does, with a 26\,per cent larger mass loss rate calculated here. P13C16U4 and P13C14U6 show a mass loss rate 35\,per cent and 28\,per cent larger than earlier, while the degree of condensation and dust-to-gas ratios are 15\,per cent smaller than before, for both models. As will be seen in Sect.~\ref{sec:discussion} P13C14U6 is very sensitive to the numerical method which partly justifies its deviation. The new P13C16U4 develops a quasi-stationary outflow (similar to that of the model using a non-adaptive grid) while the structure of the previously calculated model is more irregular. Summarising four out of six models show a larger mass loss rate here than before. Otherwise, the values are fairly similar.

A comparison between models calculated using either the VWvL or PPM advection schemes and a non-adaptive grid exhibits a very close agreement in most cases. These two volume-weighted second and third order spatially accurate schemes are implemented in different subroutines. The good agreement of the respective results indicates correct implementations.

\section{Discussion}\label{sec:discussion}
In this section results are discussed which are important for a physical interpretation of wind models. Effects of numerical origin are treated in Sect.~\ref{sec:disnumerics} -- which can be omitted by readers not concerned with details of the numerical method. Changes to the physical structure, which are obtained with the improved numerical descriptions, are discussed in Sect.~\ref{sec:disphysical}.

\subsection{Effects of numerical modifications}\label{sec:disnumerics}
The response of a stellar wind model to the different numerical adjustments introduced in Sect.~\ref{sec:numerical} is complex. For clarity, the influence of each individual modification is in the following discussed separately. Effects of the adopted advection scheme for both drift and PC models are treated first in Sect.~\ref{sec:disadv}. Thereafter differences in the physical structure between models using a non-adaptive and adaptive grid are studied in Sect.~\ref{sec:disgrid}. Followed by a discussion of the influence of the number of gridpoints in Sect.~\ref{sec:disgridpoints}.

\subsubsection{Differences due to the accuracy of advection schemes}\label{sec:disadv}
In this subsection the outcome is compared in detail of models using the three advection schemes introduced in Sect.~\ref{sec:numadv}, i.e., vL, VWvL, and PPM. First PC models are treated which use an adaptive grid; thereafter, drift and PC models using a non-adaptive grid.

In models that use an adaptive grid shocks are resolved locally by densely arranged gridpoints. A more accurate advection scheme ought not be able to improve the outcome, unless the improved accuracy is significant also in unresolved regions. Moderate differences ($<\!\!20$\,per cent; increasing, as well as decreasing) are found between properties of models calculated using either the vL or VWvL advection schemes. The shape of the physical structure is not affected -- all remain irregular. Two exceptions are P13C14U6 and P13C16U6$_{\ngrid=100}$. P13C14U6 is a highly numerically sensitive model for which all properties change significantly with the advection scheme (by \mbox{20--32}\,per cent); even the terminal velocity. The sparsely resolved model P13C16U6$_{\ngrid=100}$ shows a difference of about 30\,per cent in the dust-to-gas density ratio. It seems unnecessary to make further improvements to these models as the adaptive grid itself is found to introduce larger errors (see next subsection).

With a non-adaptive grid, which does not track shocks, shocks are not resolved by many gridpoints and the accuracy of the advection scheme is always important. If a numerical scheme is robust models using increasingly accurate advection schemes should result in more similar results. When comparing PC models using the VWvL and PPM schemes, differences in the outcome are for all presented properties $\la\!6$\,per cent, with the exception of P10C18U6 and P13C16U6$_{\ngrid=100}$. For P10C18U6 the dust-to-gas density ratio is 18\,per cent larger in the PPM model compared to the VWvL model. For the low resolution model P13C16U6$_{\ngrid=100}$ differences are 6.5\,per cent (in the terminal velocity) to 20\,per cent (in the degree of condensation and dust-to-gas density ratio). In these cases the increased accuracy is important and enhances the temperature sensitive dust formation. The third-order spatially accurate PPM models are, moreover, better at preserving the physical structure than VWvL models. A periodic structure -- although mostly of small amplitude -- is found far out both in the gas and dust, also with as few as 100 gridpoints with PPM. The same structures are more often smeared out when using less accurate advection schemes.

Corresponding comparisons between PC models using the vL and PPM advection schemes show much larger differences in outflow properties; from \mbox{1.3--110}\,per cent. For the P13C16U6 models using different number of gridpoints (Table~\ref{tab:restag}) the same changes are \mbox{1.3--30}\,per cent; being the largest for the {\ngrid}=100 model. The amount of dust is, furthermore, found to increase by \mbox{4.5--110}\,per cent for all models. The structure is finally found to be more stable with increased advection accuracy -- tending towards stationary outflows with small periodic fluctuations. These stationary outflows are not to be confused with stationary winds, where velocities stay unchanged everywhere.

For drift models using a non-adaptive grid the increased precision of the VWvL scheme, compared to vL, has a strong influence on the entire wind structure. Quantitatively, the degree of condensation is found to increase by \mbox{1.2--50}\,per cent when using the VWvL scheme, while the drift velocity changes by \mbox{3.9--57}\,per cent; for P10C18U6 there is even a change of sign. The vL scheme is not capable of calculating the correct structure, and the structure type is affected for all drift models. VWvL models in every case have a shorter periodicity, or a stationary outflow, compared to the corresponding vL model. They also, in most cases, show a smaller variability (seen in the $\hat{r}$ number of each property).

It is difficult to predict which model will be more affected by the accuracy of the advection scheme. The final conclusion is that one should use a scheme as accurate as possible, but nothing less accurate than VWvL. In particular, for drift models there is a risk that the physical structure vanishes completely in numerical noise with a less accurate advection scheme.

\subsubsection{Comparing models using a non-adaptive/adaptive grid}\label{sec:disgrid}
The change from an adaptive to a non-adaptive grid turns out to be of fundamental importance to physical properties of a stellar wind. Although an adaptive grid equation is able to resolve a small number of physical features, e.g., shocks (see Sect.~\ref{sec:numgridna}), it is not able to resolve a large number; because, there are simply not enough gridpoints. In addition, regions away from resolved features are poorly resolved. In contrast, a disadvantage with a non-adaptive grid is that no shock will be resolved by more than one or a few gridpoints (depending on the added amount of tensor viscosity and number of gridpoints), increasing the need for a more accurate advection scheme. Additionally, very narrow shocks, possibly hosting drastically different physical conditions, can also not be studied.

\begin{figure*}
\includegraphics{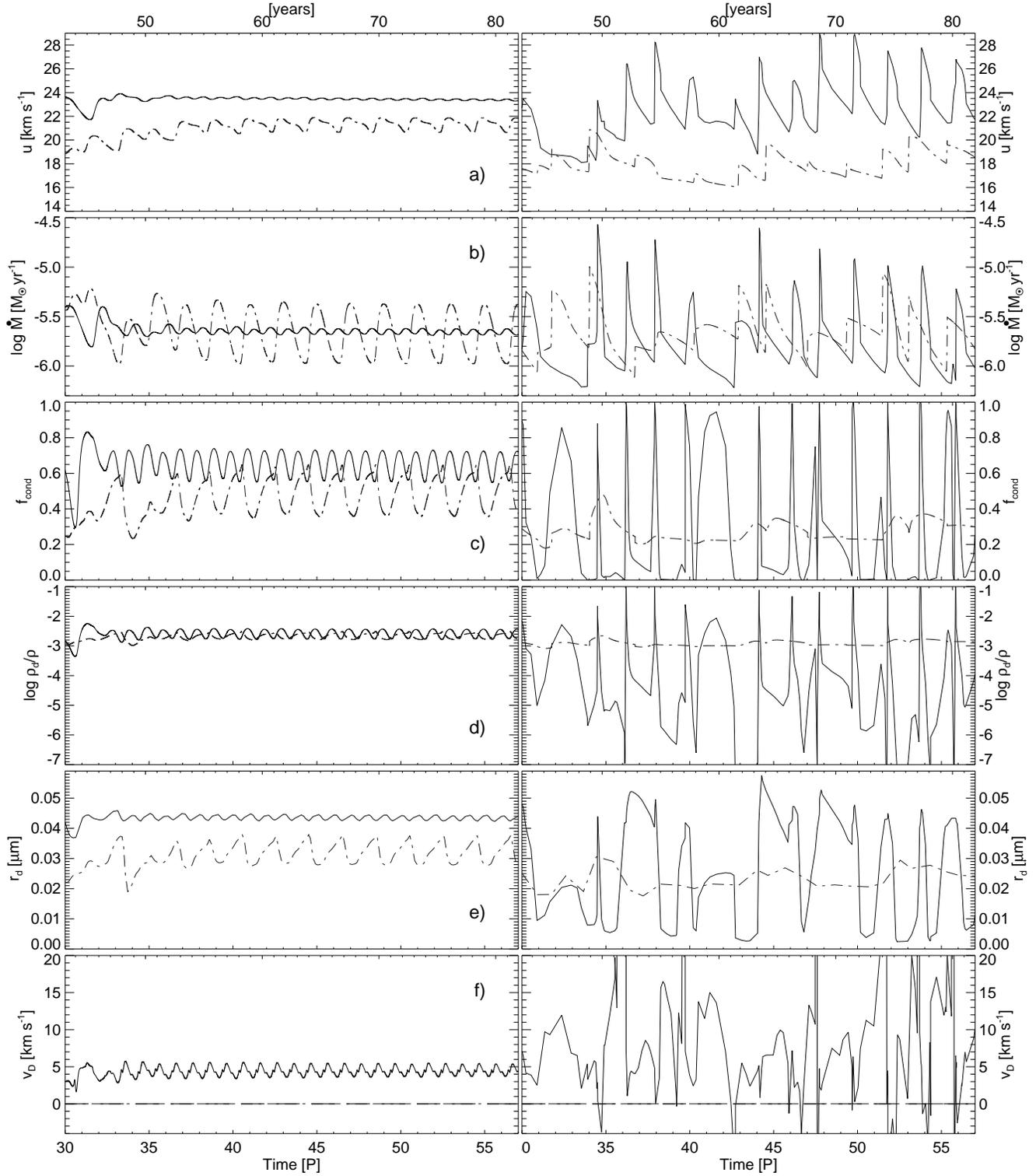}
\caption{Temporal evolution at the outer boundary, covering a time span of $25\,\pP$ (piston periods; $=\!36$ years). The model P10C18U6 is shown in the left panels using a non-adaptive grid, and in the panels on the right using an adaptive grid (with grid weights $w_{\rho,e}=1.0$). Drift models are drawn with solid lines and the corresponding PC models with dash-dotted lines, all models use the VWvL advection scheme. From the top the panels show six properties: {\bf a)} the terminal velocity {\uinf}, {\bf b)} mass loss rate $\dot{M}$ (logarithmic), {\bf c)} degree of condensation {\fcond}, {\bf d)} dust-to-gas density ratio {\drhog} (log.), {\bf e)} mean grain radius {\drad}, and {\bf f)} drift velocity \vdri. Note that the models using a non-adaptive grid (left) are both periodic, while the models using an adaptive grid are irregular (right). The significantly larger amounts of dust and larger grains formed in the drift model are clearly seen in panels {\bf c)} and {\bf e)}. The non-zero drift velocity in {\bf f)} indicates well decoupled gas and dust components. Also note that the periodicity of the drift and PC models (left) differ; $1\,\pP$ and $2\,\pP$, respectively. Furthermore, the models on the left have not settled into a periodic variation for times shorter than about $35\,\pP$. For further details, see Sect.~\ref{sec:disgrid}.}
\label{fig:disgrid_temp}
\end{figure*}

\begin{figure*}
\includegraphics{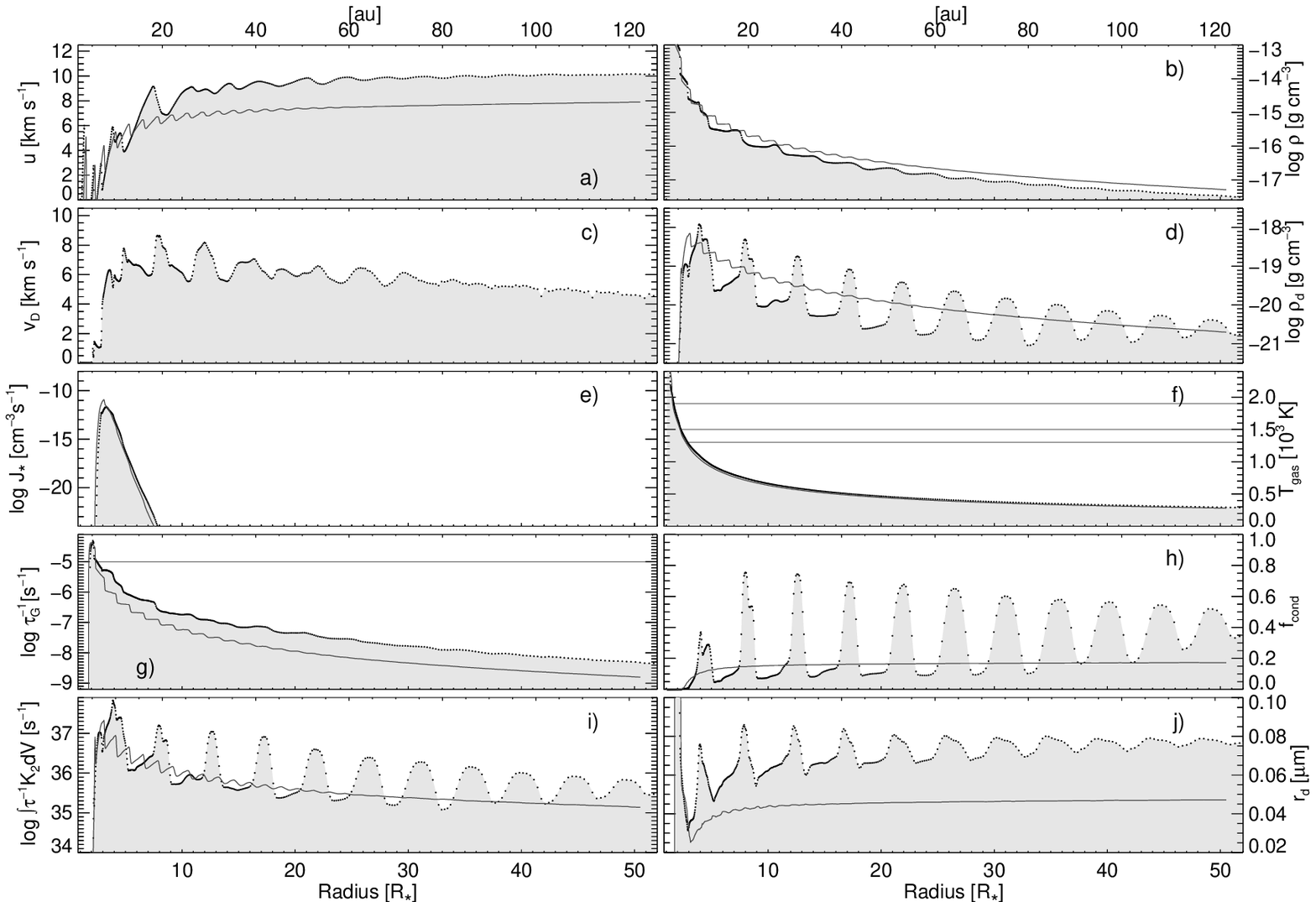}
\caption{Radial structure of an instant of the model P13C14U6 using a non-adaptive grid, illustrating the full modelled domain. The drift model is shown with a dotted filled line and the PC model (PPM) with a solid line. From the top left panels show ten properties: {\bf a)} the gas velocity $u$, {\bf b)} gas density $\rho$ (logarithmic), {\bf c)} drift velocity {\vdri}, {\bf d)} dust density {\rhod} (log.), {\bf e)} nucleation rate $J_\star$ (log.), {\bf f)} temperature {\teg}, {\bf g)} net growth rate {\taui} (log.), {\bf h)} degree of condensation {\fcond}, {\bf i)} volume integrated growth term (cf.\@ Sect.~\ref{sec:dispadust}; log.), and {\bf j)} average grain radius {\drad}. All plots are drawn as a function of the stellar radius {\Rs} (lower axis), and astronomical units (au; upper axis). Each dot on the filled contours represents an individual gridpoint. Grey horizontal lines are guides. The multi-periodicity of this model is 2\,{\pP} (drift) and 1\,{\pP} (PC). For further details see Sect.~\ref{sec:disgrid}.}
\label{fig:disgrid_radial}
\end{figure*}

The major dichotomy between physical structures formed with the two different grids is: an adaptive grid mostly forms irregular structures, while a non-adaptive grid (in the presented cases) forms periodic stationary outflows, or nearly so\footnote{All models that use a non-adaptive grid and do not form stationary outflows use the less accurate vL advection scheme.}. This holds for both drift and PC models. A comparison of the temporal outflow structure of model P10C18U6 illustrating this point is shown in Fig.~\ref{fig:disgrid_temp} using a non-adaptive (left panels) vs.\@ adaptive grid (right panels). Furthermore, the radial structure of another model, P13C14U6, is shown in Fig.~\ref{fig:disgrid_radial} using a non-adaptive grid for both the drift and PC model. Compared to a model using an adaptive grid, shocks are only strong in the inner envelope and rapidly become weaker moving outwards. Small periodic variations of both gas and dust properties, at the location of the outer boundary (originating in the piston), are seen for all quantities in Fig.~\ref{fig:disgrid_temp} (left panels). Structural shapes of the other models using a non-adaptive grid are very similar to those shown in Figs.~\ref{fig:disgrid_temp} \& \ref{fig:disgrid_radial}.

These findings strongly indicate that an adaptive grid equation mostly is uncapable of tracking many weak structures, simultaneously providing enough resolution in unresolved regions. At least not in its normally adopted setup where grid weights (for the density and energy) are set to $w_j\!=\!1.0$ (see Eq.~\ref{eq:r}). If these weights are instead replaced with $w_j\!=\!10^{-3}$ the grid is more stiff, and structures become periodic. The number or gridpoints across a shock in the inner envelope, where the density changes by about 1\,dex, is with $w_j\!=\!10^{-3}$ around 8. With $w_j\!=\!10^{-2}$ there are about 25 gridpoints across a density jump of about 1.3\,dex, although with such grid weights the dust component is no longer periodic, but irregular. 10 gridpoints are sufficient to resolve a shock of 1\,dex magnitude (cf.\@ Sect.~\ref{sec:nummethod}). With more resolved shocks inaccuracies in unresolved regions become dominating (first for the dust) and the resulting physical structure is then affected by numerical errors. This shows that if an adaptive grid is used to track shocks in a wind, grid weights should be chosen with extreme care. If grid weights are too large it is not possible to discriminate between physically periodic and irregular structures (concerning models using a constant gas opacity, see Sect.~\ref{sec:resconst}). If further resolved shocks are important for any reason, it appears necessary to use an adaptive mesh refinement instead of an adaptive grid equation. {\rW} presents stellar winds calculated using such an approach.

In this section models calculated using a non-adaptive grid (of grid type L [and U]) are next compared quantitatively with the corresponding models calculated using an adaptive grid. Only reference models are considered of the non-adaptive (L) models.

Out of six sets of model parameters three show mainly decreased values on the averaged properties from PC models; using a non-adaptive grid. The mass loss rate of P10C16U6 is 30\,per cent lower, while the terminal velocity, degree of condensation, dust-to-gas ratio, and mean grain radius are 20, 30, 30, and 10\,per cent lower, respectively. The corresponding values for P13C14U6 are $-40$, $+5$, $-4$, $-4$, and $-10$\,per cent (with the vL advection scheme these values are larger: $-60$, $-30$, $-20$, $-20$, and $-6$\,per cent). All quantities for P10C18U4 change by 1--10\,per cent. The three remaining sets of PC models show mainly increased values. The mass loss rate of P10C18U6 is 3\,per cent lower, but the terminal velocity, degree of condensation, dust-to-gas ratio, and mean grain radius increase by 20, 90, 90, and 70\,per cent, respectively. For P13C16U4 the corresponding values are 10--40\,per cent. P13C16U6 shows different changes depending on the number of gridpoints. For {\ngrid}=500 the values are: $-4$, $+7$, $+20$, $+20$, and $+20$\,per cent, respectively, while they for {\ngrid}=100 are: $-20$, $-2$, $-3$, $-5$, and $+1$\,per cent. Consequently, not only does the structure change with the use of the adaptive grid equation, but average values also change. In the presented cases by as much as 90\,per cent from a model using the numerical setup of previous papers in the series.

Most models have also been calculated using a uniform non-adaptive grid (U). Such models do not resolve shocks well and a lot more artificial viscosity is necessary to widen jumps; about five times as much is used in the central parts of the wind, typically. Nevertheless, the outcome is similar to that of the other models using a non-adaptive logarithmic grid. Differences are smaller with PC models than with drift models. Results of two such (PC and drift) models, P13C16U6, are shown in Table~\ref{tab:restaq}. Their averaged properties deviate less from the properties of the two models using a logarithmic non-adaptive grid, than they do from the properties calculated using an adaptive grid. However, since these models are not capable of reproducing results to the same accuracy as models using a logarithmic grid they are not considered here further.

\subsubsection{Importance of sufficient global spatial resolution}\label{sec:disgridpoints}
Compared to how well shocks are resolved locally, the global resolution of the full model domain also influences results notably. If a physical structure is well enough resolved, a stable numerical algorithm ought to reproduce results with an increased number of gridpoints. Exactly what happens with fewer gridpoints depends on how important narrow structures -- that are then unresolved -- are to the solution. In this section the results presented in Table~\ref{tab:restag} are discussed. To simplify the interpretation, presented PC and drift models are compared with the respective P13C16U6$_{\ngrid=500}$ reference model. Models using a non-adaptive grid are discussed first.

\begin{figure}
\includegraphics{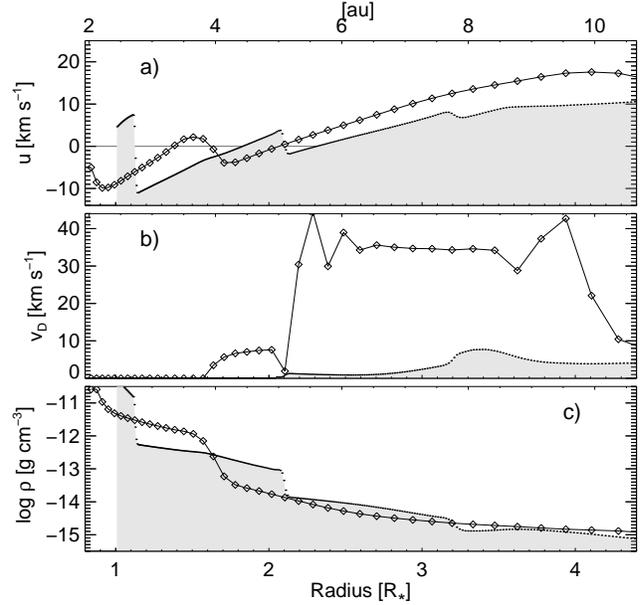}
\caption{Radial structure of an instant of the inner wind formation region of the two drift models P13C16U6$_{\ngrid=700}$ (dotted filled line) and P13C16U6$_{\ngrid=100}$ (diamonds, $\diamond$); dots and diamonds show positions of individual gridpoints. Three quantities are shown: {\bf a)} the gas velocity $u$, {\bf b)} drift velocity \vdri, and {\bf c)} gas density $\rho$ (log.). Note the wide (unresolved) shocks in the model using $\ngrid\!=\!100$. For further details see Sect.~\ref{sec:disgridpoints}.}
\label{fig:disgrires}
\end{figure}

Shocks are always resolved with at least one gridpoint on a shock in the models with $\ngrid\!=\!500$ and 700; see Fig.~\ref{fig:disgrires} (dotted filled line) for an illustration. All properties of the drift model with $\ngrid\!=\!700$ are found to differ by $\la\!4.1$\,per cent, and of the PC model by $\la0.40$\,per cent, from the corresponding reference model. The larger differences between the drift models are likely due to the lower accuracy of the VWvL advection scheme. Changes to the physical structure are negligible, albeit ``fluctuations'' are slightly larger for the model using {\ngrid}=700 (hence the difference in the characterised type in Table~\ref{tab:restag}). With $\ngrid=300$ shocks are not well resolved, but average values are still satisfactorily reproduced, both for the PC model -- $\la4.3$\,per cent (VWvL: $\la7.0$\,per cent) -- and the drift model; $\la4.4$\,per cent. Furthermore, while the PC model shows small periodic variations, the periodicity of the drift model changes from $1\,\pP$ to $2\,\pP$. With $\ngrid\!=\!100$ differences are more drastic and the PC model shows decreased values in all quantities ($\la20$\,per cent; VWvL: $\la33$\,per cent). How unresolved shocks with $\ngrid\!=\!100$ appear is illustrated with the drift model P13C16U6$_{\ngrid=100}$ in Figs.~\ref{fig:disgrires}a \& c; compare the solid line with the dotted line, which uses $\ngrid\!=\!100$. The drift velocity reaches higher values in this model (40\,{\kms}, and above), than it does when using more gridpoints and a more accurate advection scheme (than vL), see Fig.~\ref{fig:disgrires}b ($2\,\Rs\!\la\!r\!\la\!4\,\Rs$; compare with the dotted line using \ngrid=700).

Comparing the PC models using the original vL advection scheme and an adaptive grid with the reference PC model these all show an increased mass loss rate and decreased numbers on the remaining quantities. The differences are the smallest for $\ngrid=700$ (5.0--14\,per cent), intermediate for $\ngrid=300$ (10--21\,per cent), and the largest for $\ngrid=100$ (2.6--37\,per cent). The latter model compared to the other adaptive grid models also forms a stationary outflow -- since there are too few gridpoints available to resolve shocks.

\subsection{Physical properties of a dust-driven stellar wind}\label{sec:disphysical}
Previous sections have focused on the implementation and analysis of numerical improvements. In this section physical structures of stellar winds are studied, which form using these improvements. An emphasis is put on effects of gas-dust drift.

Compared to (non-drift) models that assume position coupling (PC), drift models have been found to show many differences. Effects of drift, as found previously, can be summarised as follows \citep[sect.~4 in][contains a more detailed summary]{CSa:03}:
\begin{itemize}
\item Larger variations occur in the physical structure, in particular in the dust component. Dust accumulates to locations of shocks, where densities are larger; and the gas-dust interaction stronger.
\item Dust and gas components are due to strong shocks, also present in the outer envelope, well coupled. The drift velocity typically assumes values in the range $0.1\!\le\!\vdri\!\le\!40\,\kms$; larger values occur between shocks where the density is lower.
\item Larger amounts of dust form when assuming drift-dependent dust formation; specifically grain growth is enhanced.
\item Mass loss rates are in many cases found to be lower. It is assuming drift harder to form certain wind models, which are found to form weak winds when assuming PC; weak winds are defined to be those where the terminal velocity \mbox{$\uinf\!<\!10\,\kms$}.
\end{itemize}
Moreover, most of the Planck-mean and constant-opacity (drift and PC) models are found to form an irregular (instead of a periodic) structure; although constant-opacity models are occasionally periodic. Due to larger variations in the dust component drift model structures are more irregular than those of PC models. Conclusions are sometimes ambiguous. For example, with an irregular structure it is difficult to see to what extent dust accumulates to shocks, and occasionally a drift model is found to generate a higher mass loss rate. In all cases, however, drift models form more dust.

Disentangling the interplay between a large number of physical processes (and a numerical dependence) is difficult with a strongly irregular structure. Differences are with the new more periodic models presented here found to be more clear and pronounced.

Wind structures, variation patterns, and amount of dust formed are, furthermore, highly dependent on the adopted physical assumptions and parameters describing the wind model. Absolute numbers are expected to change with the input. Time-dependency, opacities, piston properties, dust properties, and stellar parameters are such important assumptions.

In the following discussion only models using the best available advection scheme and a non-adaptive grid are used; these models are marked with ``\textsc{ref.}'' in Table~\ref{tab:restaq}. Issues of wind variability, drift velocity, and gas-dust decoupling are discussed next in Sect.~\ref{sec:dispvar}. Thereafter, revised findings on the efficiency of dust formation are treated in Sect.~\ref{sec:dispadust}, followed by a quantitative comparison of how remaining averaged properties differ between drift and PC models in Sect.~\ref{sec:dispchange}. Finally, some issues for model improvement are discussed in Sect.~\ref{sec:dispimprove}.

\subsubsection{On the drift velocity and variability of model structures}\label{sec:dispvar}

\begin{figure*}
\includegraphics{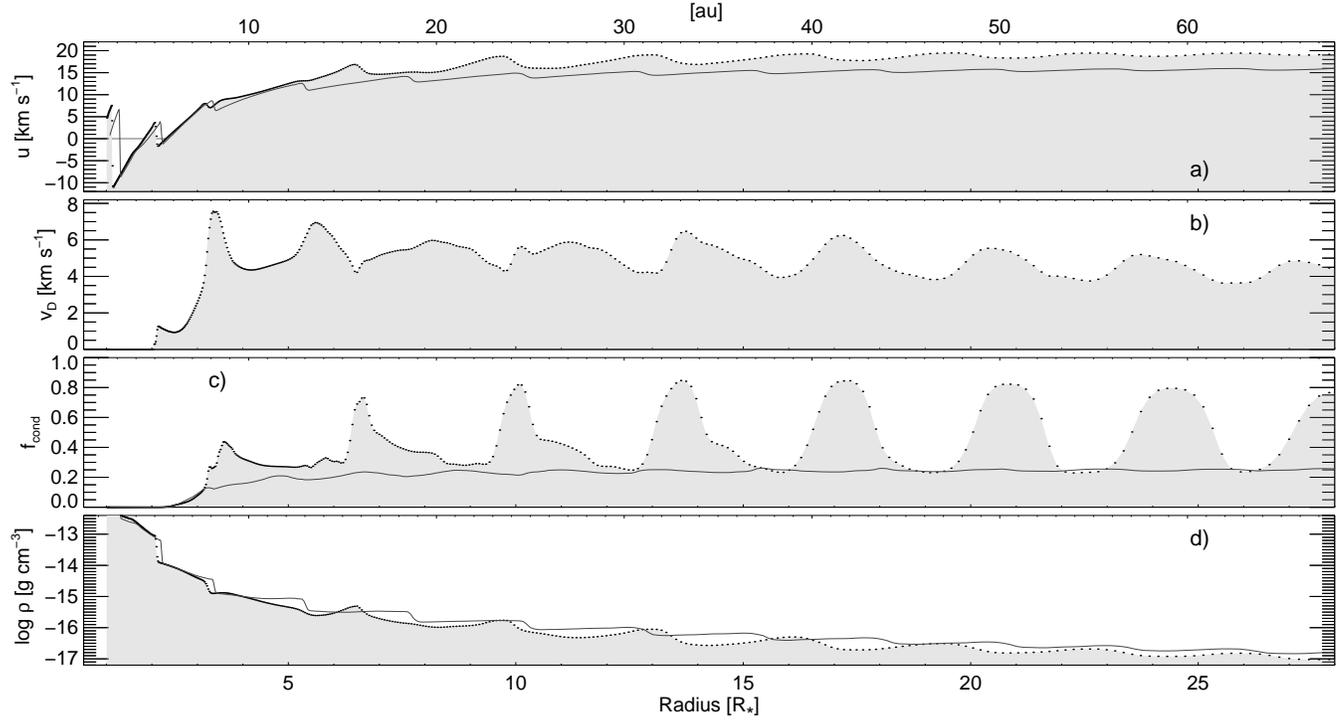}
\caption{Radial structure of the model P13C16U6 using a non-adaptive grid, illustrating the inner model domain. The drift model is shown with a dotted filled line and the PC model (PPM) with a solid line. From the top the panels show four properties: {\bf a)} the gas velocity $u$, {\bf b)} drift velocity \vdri, {\bf c)} degree of condensation {\fcond}, and {\bf d)} gas density $\rho$ (logarithmic). All plots are drawn as a function of the stellar radius {\Rs} (lower axis), with a complementing measure in astronomical units (au; upper axis). Dots on the filled contours represent individual gridpoints. Note that the gas and dust components are loosely coupled, compare the locations of maxima in the gas density, drift velocity and degree of condensation for $17\,\Rs\!\le\!r\!\le\!25\,\Rs$. For further details see Sect.~\ref{sec:dispvar}.}
\label{fig:disdecoupled}
\end{figure*}

The critical finding of Sect.~\ref{sec:disgrid} was that all (revised) wind structures are periodic. On top of such periodic structures the gas component in every case shows relatively small variations. In order to see this, compare the gas velocity and density structures in Figs.~\ref{fig:disgrid_radial}a \& b and Figs.~\ref{fig:disdecoupled}a \& d with, e.g., figs.~2a \& b in {\rSaHoc} (which illustrates a model with an irregular structure). Except in the wind acceleration region and below (i.e., for $r\!\ga\!3\,\Rs$), gas shocks are weak -- typically the change across a shock is $\la\!0.7$\,dex in the density; this value is even smaller in PC models -- and rapidly decrease in amplitude moving outwards. The dust component likewise shows very small variations in PC models; an exception is P10C18U6, which shows large variations for both components (and also strong dust formation).

Drift models show a much more variable dust component than PC models do, confirming the previous finding of variability. Unlike earlier, the dust component is now much more decoupled from the gas. A consequence partly linked with shocks, which are weaker than earlier when structures were irregular; the gas-dust interaction is therefore also weaker. Although dust (primarily) forms in the dense environment of a shock, it then drifts ahead of the gas (since $\vdri>0$). This partial decoupling is shown for model P13C16U6 in Fig.~\ref{fig:disdecoupled}. Comparing the locations of the two dust ``shells'' in the degree of condensation (Fig.~\ref{fig:disdecoupled}c) with the locations of peaks in the gas density (Fig.~\ref{fig:disdecoupled}d; and velocity, panel a) for \mbox{$16\,\Rs\!\le\!r\!\le\!25\,\Rs$}, it is seen that they do not match. Also note that the drift velocity (Fig.~\ref{fig:disdecoupled}b) in the same region attains peak values in front of the (weak) gas shocks -- sometimes displaced from the locations of dust shells. Differing locations of maxima in various gas and dust properties indicate that an interaction between the two components is still taking place. An interaction ``breaking'' the dust relative to the gas (this is seen in the decreased drift velocity with radius, Fig.~\ref{fig:disgrid_radial}c).

What is not seen in the presented figures is that the entire drift velocity structure is oscillating at a large amplitude, on top of the variations with radius seen in, e.g., Fig.~\ref{fig:disdecoupled}b. At the outer boundary the oscillations look like in Fig.~\ref{fig:disgrid_temp}f (left panel); also see the superposed periodic variations in {\rSaHob} (fig.~5 [bottom right panel]). Such oscillations demonstrate how tightly dust is bound to the radiation field.

Another new result is that the dust component in the outer envelope loses its ``patterned'' structure (i.e., dust shells are smeared out spatially); as is seen in the degree of condensation in Fig.~\ref{fig:disgrid_radial}h (for $r\ga30\,\Rs$). The same behaviour is found in the other drift models as well. With current numerical limitations it is difficult to model the region beyond about $r\!=\!50\,\Rs$ due to the locally rapidly decreasing grid resolution of the logarithmic grid. Hence, any statement on what happens to the variability pattern further out is uncertain. The same dust pattern is also diffused away when using a uniform grid, but there the grid resolution problem is even more alarming; but then in the inner envelope. It consequently appears that also the dust component of drift models ultimately form stationary outflows, at larger distances from the star.

The drift velocity is for a vast majority of circumstances (measured spatially and temporally) found to achieve moderate values; $\vdri\!\la\!10\,\kms$. Temporally averaged values of the drift velocity are shown as a function of mass loss rate in Fig.~\ref{fig:disphys}d; revealing a trend of slightly decreasing values with increasing mass loss rate. The highest values occur in the wind acceleration region \mbox{$3\,\Rs\!\la\!r\la\!10\,\Rs$}, decreasing outwards. Non-thermal sputtering, requiring drift velocities of about $40\,\kms$, does not play a role (a statement likely to change if a grain size distribution is used; cf.\ Sect.~\ref{sec:dispimprove}). Moreover, the primary molecule of importance to the grain growth process is, under these circumstances, {\CtHt} (see, e.g., {\rSaHoc}). Never is the drift velocity subsonic outwards of the dust forming region. The speed ratio ($S_{\text{D}}$; e.g., {\rSaHoa}, eq.~11), is found to take on values of $S_{\text{D}}\!\approx\!2$--3 (4 for P10C16U6) in most parts of the envelope. Referring to {\rSaHoa} (fig.~1) it is found that the relative error of the used drag coefficient ($C_{\text{D}}^{\text{LA}}$; ibid., eq.~23) is about 0.5--1.5\,per cent for these values. With the high-velocity approximation ($C_{\text{D}}^{\text{HV}}$; ibid., eq.~24) corresponding errors would be 10--20\,per cent, which is why its use is not recommended.

\subsubsection{On the amount of formed dust}\label{sec:dispadust}
Dust formation can occur where the density is high enough to allow for abundant collisions between gas (and dust) particles. A simultaneous requirement is that the temperature is low enough that dust grains do not evaporate. The resulting region where grain growth is efficient has been found to be relatively small -- about \mbox{$2\,\Rs\!\la\!r\!\la\!5\,\Rs$} \citep[in the inner envelope; e.g.,][]{GaSe:87a,CDo:92,Ho:07}.

\begin{figure*}
\includegraphics{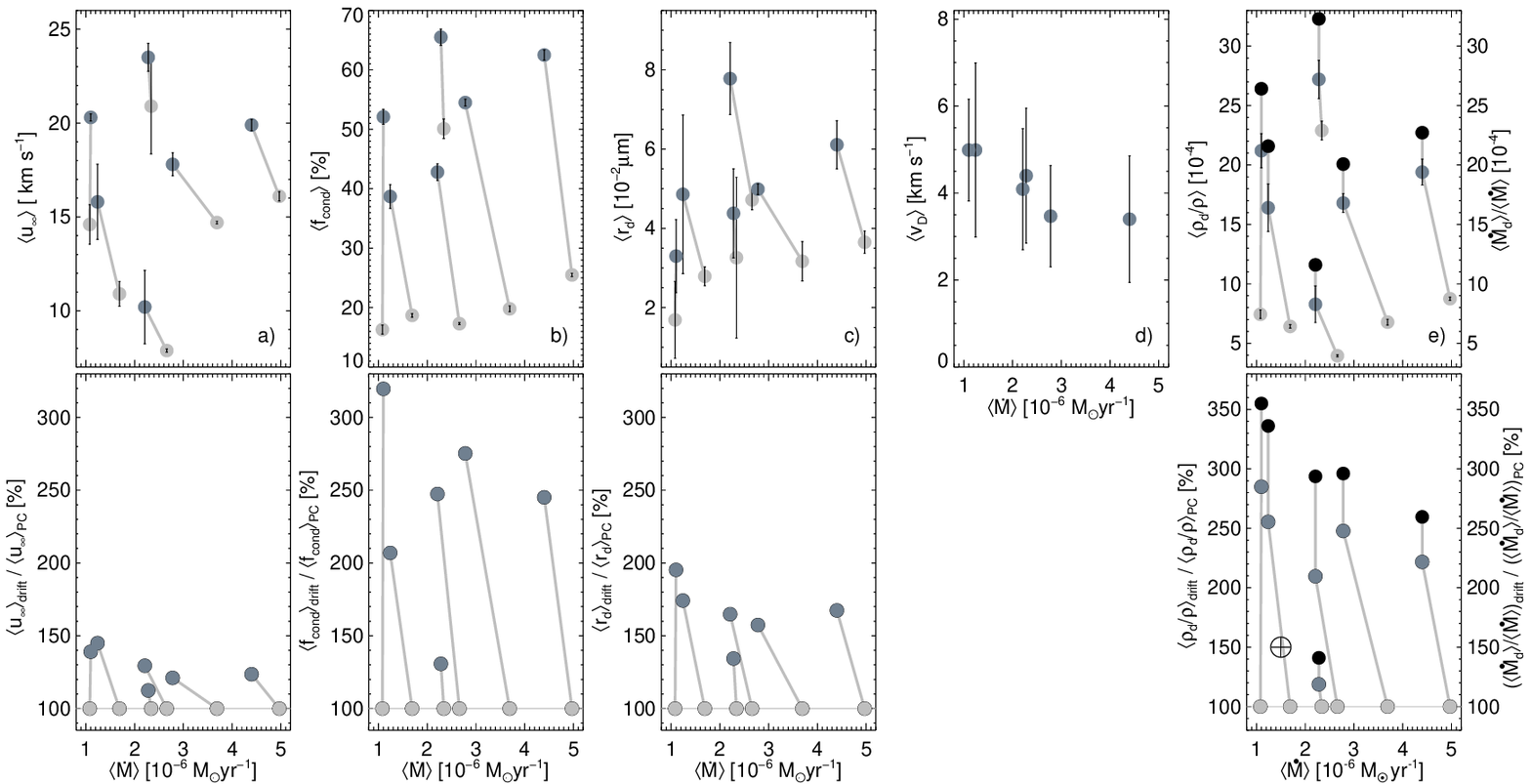}
\caption{Illustration showing how properties (from Table~\ref{tab:restaq}) of drift and PC (reference) models compare. Upper panels show absolute numbers, and lower panels the drift-to-PC ratio of the respective property (given in per cent). From the left five properties are shown: {\bf a)} terminal velocity \muinf, {\bf b)} degree of condensation \mfcond, {\bf c)} mean grain radius \mdrad, {\bf d)} drift velocity \mvdri, and {\bf e)} dust-to-gas density ratio \mdrhog (left axis) \& dust-to-gas mass loss ratio $\mdmdot/\mmdot$ (right axis); all drawn as a function of the mass loss rate \mmdot. Error bars in the upper panels show weighted relative fluctuation amplitudes (\rfluc). Moreover, drift (PC) model values are shown with dark grey (grey) bullets, \textcolor{sgrey}{$\bullet$} (\textcolor{grey}{$\bullet$}). Values of drift and PC models using the same stellar parameters are connected with grey lines. In panel {\bf e)} black bullets ($\bullet$) denote mass loss rate ratios and $\oplus$ indicates the ratio found by {\rKS} for their low mass loss model. Note that all presented models, with one exception, form drastically larger amounts of dust when assuming drift. For further details see Sect.~\ref{sec:dispadust}.}
\label{fig:disphys}
\end{figure*}

Using a stationary wind model, that included drift-dependent dust formation (and a grain size distribution), {\rKS} (see sect.~4.1) found, in one of two presented models, that the dust-to-gas mass loss ratio ($\mdmdot/\mmdot$), increased by about 50\,per cent due to drift. Such a ``flux'' ratio is more appropriate to use than the degree of condensation when comparing amounts of formed dust, since the dust component is dynamically diluted compared to the gas with a non-zero drift velocity \citep[also see][]{KrGaSe:94}.

For an averaged grain size distribution the dust-to-gas mass loss ratio is,
\begin{eqnarray}
\frac{\dot{M}_{\text{d}}}{\dot{M}}=\frac{\rho_{\text{d}}}{\rho}\frac{u+\vdri}{u},\nonumber
\end{eqnarray}
where $u$ is the gas velocity. Resulting ratios for all presented models are shown in Fig.~\ref{fig:disphys}e as a function of the mass loss rate \mmdot. As the same figure (lower panel) illustrates the increased amounts of dust due to drift, 160--250\,per cent, which are fed to the interstellar medium, are larger than the corresponding increase in the degree of condensation (110--220\,per cent); see Fig.~\ref{fig:disphys}b (lower panel). Differences between mass loss ratios and dust-to-gas density ratios, due to dilution, are about 17--40\,per cent. One model, P10C18U6, shows an efficient dust formation in the PC model (partly due to a high piston velocity and carbon-to-oxygen ratio), which is why the increase to the drift model (41\,per cent) is smaller compared to the other five models. Grain radii are, simultaneously, 34--95\,per cent larger (Fig.~\ref{fig:disphys}c).

Comparing the mass loss ratios of the bulk of the models \mbox{(2.6--3.5)} with the value of {\rKS} (\mdmdot/\mmdot$\approx$1.5), this difference is to be expected since the adopted stellar parameters are very different. Dynamical effects are less well accounted for in non-variable velocity structures of stationary winds. Features such as shocks, which promote dust formation, cannot form.

The negative slope of the dust-to-gas mass loss ratio with increasing mass loss rate seen in Fig.~\ref{fig:disphys}e (lower panel) seems to indicate that dust formation (and therefore the envelope) is less affected by drift at larger mass loss rates (and hence larger densities); this is also noticed by {\rKS}. Under such circumstances the gas-dust coupling is stronger, which leads to smaller drift velocities and supposedly more similar drift and PC models. On the contrary, at lower densities the coupling is weaker. Too few models have been calculated to permit a reliable estimate of the model behaviour outside the current sets of model parameters. It is likely that the range of mass loss rates where drift (and dust) is important is limited. This statement could be different if it turns out that models with higher mass loss rates form irregular winds (e.g., by the use of other opacities, cf.~Sect.~\ref{sec:dispimprove}).

Time scales for grain growth are longer in the outer envelope than in the inner, since densities there are lower. The volume of a shell in a radial interval increases with radius in a spherical geometry, which is why the volume-integrated amounts of formed grains still could be significant. In Fig.~\ref{fig:disgrid_radial}i the volume integrated dust growth (source) term of the moment equation corresponding to the dust (number) density is shown (i.e., the first source term on the right hand side of the $K_3$ moment equation; see, e.g., {\rSaHoc}, eq.~2). This term is a lot flatter than the net growth rate ($\tau^{-1}_{\text{G}}$, Fig.~\ref{fig:disgrid_radial}g), and decreases by less than a factor of two across the modelled envelope. Comparing the peaks of formation in the drift model the same decrease is even smaller; about a factor of one. In Fig.~\ref{fig:disgrid_radial}g the growth rate of the drift model is about 0.7\,dex larger than it is in the PC model for \mbox{$r\!\ga\!10\,\Rs$}. The grain growth source term can be integrated across the envelope for both the drift and PC model in the shown instance. Doing this it is found that 12\,per cent (PC: 8.6\,per cent) of the total amount of formed dust forms at radii $r\!\ge\!\,10\,\Rs$. It should be noted, however, that the dust formation efficiency changes during a pulsation period; although these ratios stay about the same for this model.

\subsubsection{On the change of averaged properties}\label{sec:dispchange}
The mass loss rate is in every presented case, but one, found to be slightly lower in drift models; the reason being a less than complete gas-dust coupling. Compared to the PC models the change is about $-1.9$--$27$\,per cent. Another new result is that the terminal velocity for all drift models is larger than in the corresponding PC model; by about 12--45\,per cent in the presented cases (Fig.~\ref{fig:disphys}a). The combination of a lower mass loss rate and larger terminal velocity indicates that the gas in a drift model is more tenuous than in a PC model (compare the drift and PC model gas densities in Fig.~\ref{fig:disgrid_radial}b).

The range of increased values of the terminal velocity seems to be in accordance with observed velocity distributions \citep[see, e.g.,][sect.~7.6.2 and references therein]{Ol:03}. These tend to peak towards higher values for (dust-enshrouded) IR C-Stars.

\subsubsection{Points for improvement with more detailed physics}\label{sec:dispimprove}
Drift models are now accurate enough that conclusions of topics first approached in {\rSaHoa} would benefit of a new study. Such topics include the role of specular vs.\ diffusive collisions (in the momentum and energy transfer) and the influence of different expressions of the drag force. Current models also leave a few additional issues open for further (and more difficult) improvements. Three such improvements are immediately apparent: using a grain size distribution in place of a mean, switching to frequency-dependent opacities, and using a two- or three-dimensional approach in place of the current one-dimensional. How these improvements could affect results are briefly discussed next.

Two studies, using stationary model formulations, have been carried out earlier where the influence of a grain size distribution has been treated (\citealt{DoGaSe:89}; {\rKS}). Since so much larger amounts of dust can form with current time dependent drift models, it is of high interest to see what the outcome is when appropriately considering a grain size distribution; instead of a mean. In reality grains of different size are affected by the radiation pressure from the central star differently. Large grains have a large cross section, absorb more radiation, and move at higher velocities. They could drift so fast that further growth is inhibited by non-thermal sputtering (see, e.g., {\rSaHoc}, sect.~2.2.2). Smaller grains drift slower and thereby have more time to grow efficiently. Forming a grain size distribution similar to that found in pre-solar meteorites \citep[see, e.g.,][and references therein]{BeAkCr.:05,NuWiJo.:06,An:07}, it appears that drift could be an important part of the explanation.

As was emphasised in Sects.~\ref{sec:respar} \& \ref{sec:resconst} opacities strongly influence properties of the wind structure. It is important to study how results of this paper -- that use grey molecular opacities -- compare to results calculated using frequency-dependent opacities \citep[as introduced by][]{HoGaArJo:03}; it is difficult to predict what the outcome would be. If it is found that such models are also periodic, permitting well decoupled gas and dust components, and showing fluctuations of moderate amplitude, qualitative results should show some agreement.

Larger variations and patterned structures are present in the dust component, not the gas. From the spatial distribution of dust shells in the 1D drift models presented here -- where gas is smoothly distributed -- it seems highly likely that the dust component would also be more affected than the gas in 2D and 3D models. Studies like those performed by \citet{FrHo:03} and {\rW}, where the formation of clumps can be studied, are required to show how. Those models do not, however, include any formulation for gas-dust drift currently, and will unlikely be able to reproduce an as varying dust component as found here. In this context it is worth emphasising that clumps and structures studied by {\rW} (in 2D) are calculated using a constant gas opacity, for which outflows in a majority of cases are found to be irregular (see Sect.~\ref{sec:resconst}, and ibid.\@ sect~3.1). If the understanding of how model structures form is to improve significantly it seems necessary to use 2D or 3D models, realistic opacities, and drift simultaneously.

\section{Conclusions}\label{sec:conclusions}
In recent years models of dust-driven winds from AGB stars have been made increasingly realistic with the refinement and addition of more physics. These models must handle many complex non-linear and time-dependent processes that describe a combined evolution of gas, dust, and a strong radiation field. Solving the resulting radiation hydrodynamic system of equations puts large demands on the numerical method used, which has to correctly treat, e.g., narrow shocks, and time scales covering many orders of magnitude. Results of previous studies (using the same modelling approach as in earlier papers of this series) show a majority of wind structures that are irregular (i.e., mildly chaotic). There is no thorough study where the occurrence and extent of such irregularities are clarified.

The purpose of this paper was to fill this gap by studying the influence of the numerical method on wind models that use the so-called adaptive grid equation \citep[][]{DoDr:87}. The study was based on the model description introduced in \citet[{\rSaHoa}--{\rSaHoc}]{SaHo:03,SaHo:03b,SaHo:04}. Clarifying the numerical influence all physical assumptions and parameters were kept unchanged. As before the emphasis was on improving the understanding of effects of gas-dust drift. In order to reassess previous results models were calculated both allowing drift (drift models), and not allowing drift (position coupled [PC] models).

An important result of this study is that an adaptive grid equation is not in general capable of tracking a large number of shocks without introducing numerical errors into the solution. All presented and revised models using a non-adaptive grid (which does not track shocks) were found to form periodic structures with a stationary outflow showing small amplitude fluctuations (originating in the piston). This was found to be the case both for drift and PC models. The same structures were all irregular when the grid was instead adaptive (tracking shocks). Although shocks are well resolved with an adaptive grid, regions between shocks simultaneously become unadequately resolved; because, there is only a fixed amount of gridpoints. In drift models dust is not bound to the gas and might require an appropriate resolution also between shocks. Using such a powerful tool as the adaptive grid equation without first assuring that its numerical influence does not dominate over the physical structure it is impossible to tell whether a modelled wind is physically irregular, or if variations are of numerical origin.

In attaining a high accuracy it is, moreover, necessary to use a precise advection scheme and a sufficient number of gridpoints. Detailed studies showed that smaller amounts of dust form with unresolved shocks. Incorrect structures also formed with a less accurate advection scheme. Using the PPM scheme \citep{CoWo:84}, numerical errors were found to be small; at the per cent level. Compared to previously calculated models, differences in individual model properties were overall as large as 100\,per cent.

It is important to point out that absolute values of results in this study are highly dependent on adopted physical assumptions and parameters, as well as the range of stellar parameters used. Qualitatively, the results are comparable with, or applicable to, models that also turn out to form periodic structures. An application to models which instead form irregular structures is not meaningful, but requires a separate study.

Many physical results of previous papers in this series were confirmed, but there were also new findings. The gas component typically showed small variations. Unlike earlier there were no shocks in the outer envelope. The dust component, however, showed large variations in drift models, confirming previous results. Although the periodic variation pattern diffused away at larger radii; and it consequently appears that also the outflow of the dust component becomes stationary. Moreover, unlike previous results the dust component was less tightly coupled to the gas. Typically the drift velocity was fairly low in the entire envelope, $\vdri\!\la\!10\,\kms$. The highest values were assumed in the inner envelope, decreasing outwards to $3\!\la\!\vdri\!\la\!5\,\kms$ at the outer boundary (at about 50 stellar radii).

Drastically larger amounts of dust formed in drift models compared to PC models; the increase was 160--250\,per cent in every case but one (for which the increase was 40\,per cent). It was, furthermore, found that mass loss rates typically were lower in drift models, by $\la30\,$per cent. Terminal velocities were, to the contrary, larger, by 10--50\,per cent. Winds of drift models were consequently more tenuous than those of PC models. Altogether results (again) show that neglecting gas-dust drift creates imprecise outcome.

The modelled winds were neither the least, nor the most massive. Partly this is a consequence of the Planck mean absorption coefficients used, which result in less dense wind structures, than would be the case had instead frequency-dependent opacities been used. Partly because weak winds are difficult to calculate with drift. It is highly likely that the range of mass loss rates where drift (and dust) is important is limited. Finally, there are issues for model improvement where a closer study would make the understanding of the wind formation mechanism more complete. Three such issues addressed were: using a grain size distribution instead of a mean size, using frequency-dependent opacities, and the role of a two- or three-dimensional modelling approach.

\section*{Acknowledgements}
I thank Marina Skender and Hakan \"Onel for carefully reading the manuscript and providing useful comments for its improvement.

\bibliographystyle{aa}
\bibliography{CS_Refs}

\label{lastpage}
\end{document}